\documentclass[prd,twocolumn,notitlepage,showpacs,preprintnumbers,amsmath,amssymb,nofootinbib,aps,10pt,longbibliography,superscriptaddress]{revtex4-2}

\pdfoutput = 1

\usepackage{dcolumn}
\usepackage{bm}
\usepackage{xcolor}
\usepackage[utf8]{inputenc}
\usepackage[spanish,english]{babel}
\usepackage{amsmath,amssymb,amsfonts,latexsym,cancel}
\usepackage[normalem]{ulem}
\usepackage{graphicx}
\usepackage{color}
\usepackage{soul}
\usepackage{ulem}
\usepackage[colorlinks=true,linkcolor=red,urlcolor=blue,citecolor=blue]{hyperref}
\usepackage{amsmath}
\usepackage{slashed}
\usepackage{braket}
\usepackage{amssymb}
\usepackage{amsmath}

\usepackage{orcidlink,booktabs}
\usepackage{siunitx,bm}
\usepackage{comment}

\usepackage{graphicx}
\allowdisplaybreaks

\newcommand{\mi}{\mathrm{i}}
\renewcommand{\vec}[1]{\boldsymbol{\mathbf{#1}}}

\begin{document}
\preprint{CALT-TH/2025-007, N3AS-24-040} 
\title{Detecting gravitational signatures of dark matter with atom gradiometers  }

\author{Leonardo Badurina}
\email{badurina@caltech.edu}
\affiliation{Walter Burke Institute for Theoretical Physics, California Institute of Technology, Pasadena, CA 91125, USA}

\author{Yufeng Du}
\email{yfdu@caltech.edu}
\affiliation{Walter Burke Institute for Theoretical Physics, California Institute of Technology, Pasadena, CA 91125, USA}

\author{Vincent S. H. Lee}
\email{vincentszehimlee@berkeley.edu}
\affiliation{Walter Burke Institute for Theoretical Physics, California Institute of Technology, Pasadena, CA 91125, USA}
\affiliation{Department of Physics, University of California Berkeley, Berkeley, CA 94720, USA}
\affiliation{Department of Physics, University of California, San Diego, La Jolla, CA 92093-0319, USA}

\author{Yikun Wang}
\email{yikunw@jhu.edu}
\affiliation{Walter Burke Institute for Theoretical Physics, California Institute of Technology, Pasadena, CA 91125, USA}
\affiliation{Department of Physics \& Astronomy, The Johns Hopkins University, Baltimore, MD 21218, USA}

\author{Kathryn M. Zurek}
\email{kzurek@caltech.edu}
\affiliation{Walter Burke Institute for Theoretical Physics, California Institute of Technology, Pasadena, CA 91125, USA}
\date{\today}

\begin{abstract}

We study the purely gravitational signatures of dark matter from the ultralight to the ultraheavy mass range in proposed long-baseline atom gradiometers, focusing on terrestrial designs, such as AION-km and MAGIS-km, as well as space-based concepts, such as MAGIS-space, AEDGE and AEDGE+. Due to its exceptional acceleration sensitivity and depending on astrophysical backgrounds, a detector similar to AEDGE+ could detect a dark matter subcomponent which constitutes $\mathcal{O}(10\%)$ of the local dark matter energy density and is populated by compact clumps of mass between $10^6$~kg and $10^{10}$~kg ($10^{-25}~M_\odot\lesssim M \lesssim 10^{-21}~M_\odot$) in an otherwise unexplored region of dark matter model space. 
Furthermore, because the gravitational observable depends on the relative gravitational time delay measured by spatially separated atomic clouds, we find that atom gradiometers are parametrically more sensitive than laser interferometers, such as LIGO and LISA, to fast-oscillating spacetime perturbations sourced by energy density and pressure fluctuations of ultralight dark matter. Depending on astrophysical backgrounds, 
a detector akin to AEDGE+ could probe a DM overdensity of $\mathcal{O}(10)$ times the local dark matter energy density for masses $m\lesssim 10^{-17}$~eV.
\end{abstract}

\maketitle

\section{Introduction}
\label{introduction}

Dark matter (DM) remains one of the most important and unsolved problems in fundamental physics. Assuming a single species, its mass could range from approximately $10^{-21}$~eV~\cite{Zimmermann:2024xvd} to asteroid masses~\cite{Green:2024bam}, and its existence is currently inferred exclusively through gravitational interactions with Standard Model particles on galactic, astrophysical and cosmological scales.

Because of DM's crucial influence on the evolution of our Universe, it is imperative to explore the vast phenomenological landscape of models and interactions. Indeed, significant effort has been dedicated to the detection of DM in the lab across a wide range of masses, spins and couplings. Nevertheless, it is natural to consider whether any DM candidate may be detected \textit{exclusively} via gravitational interactions with a laboratory probe.~For example, compact clumps of DM would induce transient distortions of spacetime, giving rise to transient accelerations on test masses and corrections to the propagation of photons, and therefore to measurable effects in  accelerometers~\cite{Carney:2019pza,Blanco:2021yiy,Windchime:2022whs} and laser interferometers~\cite{Adams:2004pk,Seto:2004zu,Hall:2016usm, Kawasaki:2018xak, Jaeckel:2020mqa, Baum:2022duc, Lee:2022tsw,Du:2023dhk}. These clumps, which would also give rise to observable effects in pulsar timing arrays~\cite{Siegel:2007fz,Ramani:2020hdo,Dror:2019twh,Clark:2015sha,Kashiyama:2018gsh,Schutz:2016khr,Baghram:2011is,Lee:2020wfn,Lee:2021zqw}, could be primordial black holes~\cite{Carr:2016drx,Carr:2020xqk,Chakraborty:2022mwu}, or composite states of non-baryonic matter~\cite{Wise:2014ola,Wise:2014jva,Hardy:2014mqa,Gresham:2017cvl,Gresham:2017zqi,Bai:2018dxf}, provided that the latter are sufficiently dense and stable. At much smaller masses, ultralight dark matter (ULDM) candidates would also give rise to purely gravitational signatures in the lab. Aside from being potentially observable in pulsar timing arrays~\cite{Khmelnitsky:2013lxt, Kim:2023kyy,Boddy:2025oxn} and precision astrometry~\cite{Kim:2024xcr,Dror:2024con}, ULDM energy density fluctuations would induce time-dependent spacetime fluctuations, and therefore to observable effects in laser interferometers~\cite{Kim:2023pkx}.

Aside from these probes, it is natural to ponder whether ground-breaking quantum sensing techniques, which are expected to become key in the push towards the precision frontier (see Ref.~\cite{Safronova:2017xyt} for a comprehensive review), could provide the prospect of detecting DM via purely gravitational interactions. 

Atom interferometers (AIs) are versatile quantum sensors that can be employed in different configurations for a wide variety of precision measurements. In particular, atom gradiometers (AGs), which consist of two spatially separated AIs that are referenced by common lasers, are poised to play a key role in uncovering the nature of DM. For example, future large-scale terrestrial AG experiments (see Refs.~\cite{Proceedings:2023mkp,Proceedings:2024foy} for recent reviews), such as AION~\cite{Badurina:2019hst}, MAGIS~\cite{MAGIS-100:2021etm}, MIGA~\cite{Canuel:2017rrp}, ELGAR~\cite{Canuel:2019abg}, and ZAIGA~\cite{Zhan:2019quq}, and proposed space-based experiments such as MAGIS-space~\cite{Graham:2017pmn}, STE-QUEST~\cite{STE-QUEST:2022eww}, AEDGE~\cite{AEDGE:2019nxb} and AEDGE+~\cite{Badurina:2021rgt}, the latter being based on the AGIS proposal~\cite{Dimopoulos:2008sv}, will become powerful probes of ULDM with masses between $10^{-16}$~eV and $10^{-12}$~eV. Scalar ULDM with dilatonic couplings to Standard Model operators would give rise to time-varying oscillations in atomic transition frequencies~\cite{Stadnik:2015kia}, and therefore to an observable signal in AGs~\cite{Arvanitaki:2016fyj, Badurina:2021lwr, Badurina:2022ngn, Badurina:2023wpk}.~In light of their exquisite sensitivity to accelerations, co-located interferometers operating with atomic species with different neutron number would also be sensitive to the time-varying force mediated by a $U(1)_{B-L}$ gauge boson~\cite{Graham:2015ifn}. Furthermore, via collisional decoherence, it may be possible to constrain sub-GeV DM via spin-independent DM-nucleon interactions~\cite{Du:2022ceh,Badurina:2024nge}. Recently, it has been suggested that AGs could also be used to constrain Lorentz-invariant and Lorentz-violating models of spin-2 ULDM \cite{Blas:2024jyh}.

Proposed long-baseline AGs are also poised to become the best probe of gravitational waves in the mid-frequency band (i.e. between $10^{-3}$~Hz and 1 Hz). It is expected that these experiments could probe gravitational waves originating from exotic compact objects~\cite{Banks:2023eym}, intermediate-mass black holes, phase transitions in the early Universe and networks of cosmic strings~\cite{Dimopoulos:2006nk,Dimopoulos:2007cj,Graham:2012sy,Graham:2016plp,Graham:2017lmg,Ellis:2020lxl,Badurina:2021rgt,Baum:2023rwc}. This follows from the fact that the AG observable depends on two independent timescales: the photon roundtrip time, which depends on the spatial separation between the two AIs in an AG, and the interferometric time, which depends on the duration of a single interferometric cycle. The latter also approximately sets the peak response of a detector. Therefore, by choosing interferometric sequences with a long baseline and long interferometric time, an AG can effectively close the gap between LISA and LIGO.

In light of their projected sensitivity to weak gravitational signals in an unexplored frequency band, could proposed atom gradiometers be used to detect DM via gravitational interactions? If so, what DM masses could be detected or ruled out? To address these questions, we take a bottom-up approach: we first study the purely gravitational phase shift induced by transient and compact clumps of DM, whose size is negligible relative to all experimental length scales; subsequently, we explore the phase shift induced by energy density and pressure fluctuations of ultralight dark matter. To compute these effects, we utilize the framework developed in Ref.~\cite{Badurina:2024rpp}, which decomposes the gravitational AG phase shift into Doppler (i.e. tidal displacement of the atoms along the baseline), Einstein (i.e. gravitational redshift or time delay measured by the atoms), and Shapiro (i.e. time delay accrued by the propagation of photons along the baseline) contributions, similarly to the decomposition of the observable in laser interferometers~\cite{Rakhmanov:2004eh,Lee:2024oxo} 
and pulsar timing arrays~\cite{Maggiore:2007ulw}.

Notably, we find that, depending on astrophysical backgrounds, a space-based detector akin to AEDGE+, which was not considered in Ref.~\cite{Baum:2022duc}, could detect the purely gravitational and transient signal induced by an $\mathcal{O}(0.1)$ fraction of DM populated by compact clumps in the unexplored $10^6~\mathrm{kg}\lesssim M \lesssim 10^{10}~\mathrm{kg}$ (
$10^{-25}~M_\odot\lesssim M \lesssim 10^{-21}~M_\odot$) mass window. Furthermore, because they can access the relative gravitational redshift induced by such fast-oscillating fluctuations, AGs are found to be parametrically more sensitive than laser interferometers with comparable strain sensitivity to metric perturbations sourced by the fast-oscillating fluctuations in the energy density and pressure of ULDM.
Depending on astrophysical backgrounds, this enhanced sensitivity could allow a detector like AEDGE+ to probe a DM overdensity of $\mathcal{O}(10)$ times the local DM energy density for masses $\lesssim 10^{-17}$~eV.

This paper is structured as follows. In section~\ref{sec:AG}, we review the basics of atom gradiometry and provide a brief summary of the general relativistic framework developed in Ref.~\cite{Badurina:2024rpp} for computing phase shifts. In section~\ref{sec:ultraheavy} we investigate the discovery potential of terrestrial and space-based gradiometers to dark clumps, while in section~\ref{sec:ultralight} we study the sensitivity of these experiments to energy density and pressure fluctuations of ULDM. We summarise our results in section~\ref{sec:discussion_conclusion}. Appendices~\ref{app:noise_sources}--\ref{appendix:einstein_eqn} support the discussions and calculations in sections~\ref{sec:AG}--\ref{sec:ultralight}.  
 
\section{Review of atom gradiometry} \label{sec:AG}

\subsection{Description of the interferometric sequence}

Atom interferometers (AIs) measure the phase between coherent, spatially delocalised quantum superpositions of atomic wavepackets. This can be achieved by constructing a two-level system, composed of an excited state and a ground state with energy splitting $\omega_a$. In single--photon AIs, individual laser pulses control both the transitions from ground to excited state, and vice versa, and the motion of the wavepackets' centre--of--mass via Rabi oscillations~\cite{Abend:2020djo}.
A laser pulse that interacts with the atom over a quarter of a Rabi cycle (i.e. a $\pi/2$-pulse) takes an atom in one state to an equal superposition of two states with relative velocity fixed by the laser wavevector; hence, such a pulse acts as a ``beamsplitter". A pulse over half of a Rabi cycle (i.e. a $\pi$-pulse) reverses the state of the atom and, depending on the initial state of the atomic wavepacket, either adds or subtracts momentum from the atom's centre-of-mass; thus, such a pulse acts as a ``mirror"~\cite{Foot2005}.

Atom gradiometers (AGs) are sets of spatially separated AIs operated by common and counter-propagating laser pulses, which are emitted by sources on opposite ends of a baseline of length $L$. These configurations enable differential phase shift measurements that are largely insensitive to laser phase noise~\cite{Yu:2010ss,Graham:2012sy}, thereby enabling these devices to operate near the quantum shot-noise limit, as recently demonstrated in Refs.~\cite{Rudolph:2019vcv,AION:2025igp}. 

\begin{figure}
    \centering
    \includegraphics[width=0.99\linewidth]{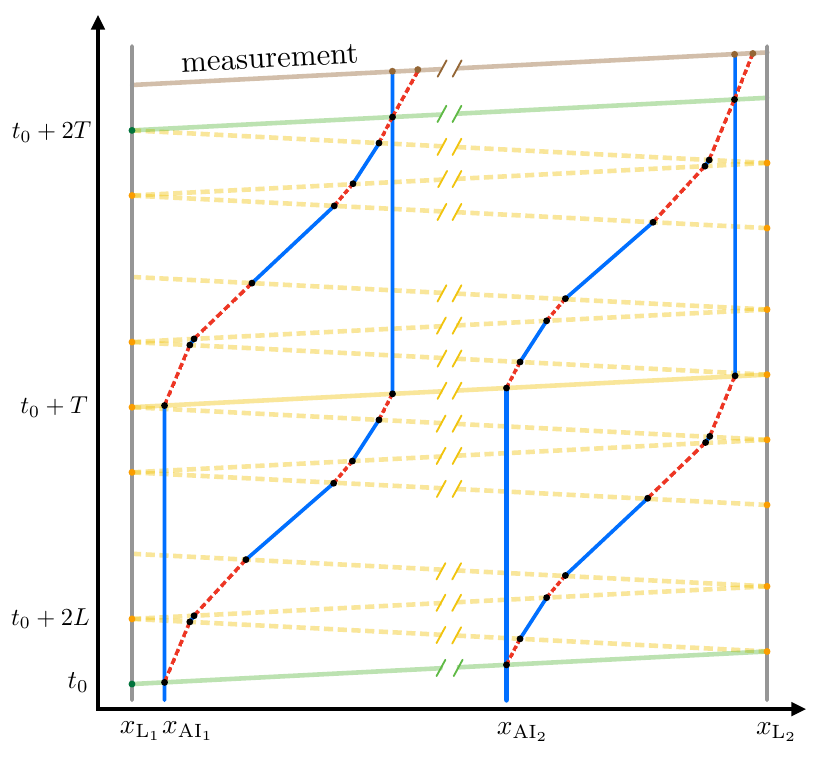}
    \caption{Spacetime diagram of a single-photon LMT atom gradiometer with $n=4$. Coordinate time and position are displayed along the vertical and horizontal axes, respectively. The laser worldlines are shown with grey solid lines.  The worldlines of the wavepackets' centre-of-mass are shown with solid blue and dashed red lines for ground and excited states, respectively. Beam-splitter ($\pi/2$) pulses are shown in green, while mirror ($\pi$) pulses are shown in yellow; mirror pulses that affect one (two) arm(s) of an AI are shown with dashed (solid) lines.  Atom-light interactions are marked with black dots. The measurement pulse is shown with a brown line.}
    \label{fig:LMT_spacetime}
\end{figure}

Since the leading order phase shift scales with the spacetime area enclosed by an interferometer~\cite{Kasevich:1991zz}, it is advantageous to use multiple photon kicks, commonly referred to as large-momentum-transfer (LMT) kicks, to increase the spatial separation between the two wavepackets in an AI, i.e. the distance between the two ``arms'' of the interferometer~\cite{McGuirk:2000zz}. Adopting the convention of Ref.~\cite{Graham:2016plp}, a configuration with $n$ LMT kicks consists of a total of $4n-1$ laser pulses: two $\pi/2$-pulses and $4n-3$ $\pi$-pulses. In Fig.~\ref{fig:LMT_spacetime} we illustrate an example of such a gradiometer configuration for $n=4$. The two $\pi/2$ (i.e., beam-splitter) pulses, which are shown with green solid lines, are applied at the beginning and at the end of the interferometric cycle, i.e., we consider a $\pi/2 - \pi - \cdots - \pi -\pi/2$ scheme. The initial and final stages of the interferometric cycle, known as the ``beamsplitter sequence'', contain $n-1$ $\pi$-pulses each, which we display with dotted yellow lines. These pulses are tuned to interact with only one arm of each AI. Depending on the internal state of the atom, these pulses add or remove momentum to the wavepacket's center-of-mass along the direction of the baseline as a result of momentum conservation. For example, after interacting with a laser pulse, a wavepacket in the ground state, whose worldline we display with a solid blue line, transitions to the excited state and recoils in the direction parallel to the laser's wavevector; instead, after interacting with a laser pulse, a wavepacket in the excited state, whose worldline we display with a dashed red line, transitions to the ground state via stimulated emission and hence recoils in the direction opposite to the laser's wavevector. 
At time $T$, which is referred to as the interrogation time and approximately marks half of the cycle's duration, $2n-1$ $\pi$-pulses are used to redirect the atomic wavepackets, thus forming the ``mirror sequence''. In this case, one of these pulses, shown with a solid yellow line, interacts with both arms of each AI. 

In this work, we specialise to single-photon AGs employing LMT. A large number of LMT kicks can be achieved by using a two-level system with a long-lived excited state, such as the clock transition in $^{87}\mathrm{Sr}$~\cite{Loriani:2018qej}: the ground and excited states correspond to $5\mathrm{s}^2\, ^{1}\mathrm{S}_0$ and $5\mathrm{s}5\mathrm{p}\, ^{3}\mathrm{P}_0$, respectively, $\omega_a\approx2.697\times10^{15}~\mathrm{rad}/\mathrm{s}$ and the natural linewidth of the transition is $\approx 10^{-3}$~Hz~\cite{Boyd_2007,Muniz_2021}. Furthermore, we consider a terrestrial experiment and two space-based proposals. For the Earth-bound detector, we consider a $1$~km baseline with two AIs at either end of the baseline. The chosen interferometric parameters allow the atomic wavepackets to remain within the extent of the baseline, as discussed in Ref.~\cite{Badurina:2022ngn}. This design, which is consistent with the AION-km and MAGIS-km proposals, is a broadband detector with optimized reach to mid-frequency gravitational waves and ultralight dark matter. For space-based proposals, we consider two distinct and ambitious configurations. In both cases, two satellites in mid-Earth orbit house the laser optics system used for manipulating and measuring the atomic wavepackets, and each AI is operated in broadband mode. The difference between the two designs lies in the spatial extent of the interferometric region: in one case, the atoms are confined within the satellite; in the other case, the atoms are allowed to evolve in the spacecraft's local environment. Consequently, the lattermost design can accommodate much larger spatial separations between the two arms of a single AI, i.e. a much longer interrogation time and a larger number of LMT kicks. Importantly, the design with a much smaller interferometric region, denoted as ``space-based (inside)" in Table~\ref{tab:exps}, is based on the MAGIS-space and AEDGE proposals. The parameters ensure that the wavepackets can remain in free-fall within a region ${\sim} 1$~m in length. The other configuration, which we define as ``space-based (outside)", is based on the AEDGE+ proposal, which is the single-photon version of the AGIS proposal.  
Importantly, in the angular frequency range set by $\omega_\mathrm{min}$ and $\omega_\mathrm{max}$, we assume that the gradiometer is shot-noise limited. In every realization, $\omega_\mathrm{max}$ is set by the repetition rate of the experiment. Due to the noise sources considered in this work, $\omega_\mathrm{min} \gg 2\pi/T_\mathrm{int}$, where $T_\mathrm{int}$ is the integration time (i.e. the duration of the measurement campaign). For the terrestrial experiment, $\omega_\mathrm{min}$ is based on estimates of seismic~\cite{Badurina:2022ngn} and atmospheric~\cite{Carlton:2024lqy} gravity gradient noise (GGN). The smallest (optimistic) and largest (benchmark) values of $\omega_\mathrm{min}$ assume significant and minimal noise mitigation via, e.g., passive or active noise mitigation strategies, respectively. For both space-based proposals, the largest value of $\omega_\mathrm{min}$ is set by the largest frequency at which the instrumental noise budget is anticipated to be shot-noise limited~\cite{Dimopoulos:2008sv}, and the smallest value of $\omega_\mathrm{min}$ is estimated from asteroid GGN~\cite{Fedderke:2020yfy}. Note, however, that astrophysical foregrounds due to, e.g., unresolved galactic and
extragalactic binary mergers~\cite{Karnesis:2021tsh,Farmer:2003pa,Staelens:2023xjn,Hofman:2024xar,Boileau:2025jkv,Lewicki:2021kmu}, may dominate the ``space-based (outside)" proposal's noise budget between $\sim 10^{-6}$~Hz and $\sim 10^{-2}$~Hz (cf. Appendix~\ref{app:noise_sources} for a detailed discussion). For simplicity and in light of existing uncertainties, we do not consider these backgrounds in our projections, but we comment on the impact of astrophysical foregrounds on our results in sections~\ref{sec:ultraheavy:sensitivity} and~\ref{sec:ultralight:sensitivity}.

\begingroup
\setlength{\tabcolsep}{10pt}
\renewcommand{\arraystretch}{1.5}
\begin{table*}
    \begin{tabular}{c c c c c c c }
		\hline
        \hline
		Design & $L$ [m] & $n$ & $T$ [s] & $\sqrt{S_n}\  [1/\sqrt{\mathrm{Hz}}]$ & $\omega_\mathrm{min}/2\pi$ [Hz] & $\omega_\mathrm{max}/2\pi$ [Hz] \\
		\hline
        km-baseline & $10^3$ & $2500$ & $1.7$  & $10^{-5}$ & $\{10^{-4}$, $10^{-1}\}$ & 10 \\
        Space-based (inside) & $4.4\times 10^7$ & $2$ & $100$ & $10^{-4}$ & $\{10^{-6}$, $10^{-3}\}$ & 10 \\
        Space-based (outside) & $4.4\times 10^7$ & $250$ & $150$ & $10^{-4}$ & $\{10^{-6}$, $10^{-3}\}$ & 10 \\
        \hline
        \hline
	\end{tabular}
	\caption{Parameters of the AG experiments considered in this work. The three designs are as follows: a  terrestrial experiment with a km-long baseline [``km-baseline"]~\cite{Badurina:2022ngn}, whose parameters are compatible with those of the MAGIS-km~\cite{Abe:2021ksx} and AION-km proposals~\cite{Badurina:2019hst}; a space-based experiment, which is based on the MAGIS-space~\cite{Abe:2021ksx,Graham:2017pmn} and AEDGE proposals~\cite{AEDGE:2019nxb}, consisting of two AIs that are confined within two satellites in mid-Earth orbit [``space-based (inside)"]; and a space-based gradiometer that enables the free evolution of the atoms outside each satellite [``space-based (outside)"], which is akin to the AEDGE+ proposal~\cite{Badurina:2021rgt} (the single-photon version of AGIS~\cite{Dimopoulos:2008sv}). Here, $L$ is the length of the baseline (which is a good approximation of the gradiometer length), $n$ is the number of LMT kicks, $T$ is the interrogation time, and $S_n$ is the shot-noise limited power spectral density. We assume that the experiments are shot-noise limited between $\omega_\mathrm{min}$ and $\omega_\mathrm{max}$.}  
	\label{tab:exps}
\end{table*}
\endgroup

\subsection{Coordinate-invariant phase shift calculation}\label{subsec:formalism}

To compute leading-order general relativistic phase shifts in an atom gradiometer, we use the simplified general relativistic framework introduced in Ref.~\cite{Badurina:2024rpp}, which is valid to leading order in the metric perturbation $h_{\mu\nu} \ll 1$ and the atom's recoil velocity.
In this framework, the observable (i.e. the difference between the phase shifts recorded by each AI) is decomposed as a sum of three distinct (and separately not diffeomorphism-invariant) contributions: the phase shift caused by the tidal displacement of the atoms along the baseline (i.e. the Doppler phase shift), the phase shift due to the gravitational redshift measured by the atoms (i.e. the Einstein phase shift) and the phase shift due to the delay in the arrival time of photons at atom-light interaction points (i.e. the Shapiro phase shift).\footnote{Note that there exist alternative methods for computing general relativistic effects in AIs, e.g., Refs.~\cite{Dimopoulos:2008hx,Roura:2018cfg,Werner:2023qfu, Roura:2024rmm}.} Focusing on LMT configurations, and quoting results from Ref.~\cite{Badurina:2024rpp}, the Fourier transform (FT)\footnote{We use the 
conventions $\widetilde{x}(\omega)=\int_{-\infty}^{\infty}dt\,x(t)e^{-\mi\omega t}$ and $x(t)=(2\pi)^{-1}\int_{-\infty}^{\infty}d\omega \,\widetilde{x}(\omega)e^{\mi\omega t}$, where $\omega$ is the angular frequency~\cite{Moore:2014lga}.} of the leading order total gradiometer observable is given by 
\begin{widetext}
\begin{equation} \label{eq:total}
\Delta \widetilde{\phi}_\mathrm{grad}^\mathrm{LMT} = \Delta \widetilde{\phi}_{\mathrm{grad},\mathcal{D}}^\mathrm{LMT}  + \Delta \widetilde{\phi}_{\mathrm{grad},\mathcal{E}}^\mathrm{LMT} 
+ \Delta \widetilde{\phi}_{\mathrm{grad},\mathcal{S}}^\mathrm{LMT} \, ,
\end{equation}
where the Doppler ($\mathcal{D}$), Einstein ($\mathcal{E}$) and Shapiro ($\mathcal{S}$) contributions can be expressed as
\begin{equation} \label{eq:LMT_tot_F}
\begin{aligned}
\Delta \widetilde{\phi}^\mathrm{LMT}_{\mathrm{grad},\mathcal{D}} (\omega) & = \frac{1}{2} k_{\rm eff} \, T^2 \,  \omega^2 \, 
 \left [n_i \delta \widetilde{x}^i (\omega,x_{\mathrm{AI}}) \left (K_+ (\omega)   e^{\mi \omega (x_{\mathrm{AI}}-x_{\mathrm{L}_1})} +  K_- (\omega) e^{\mi \omega (x_{\mathrm{L}_2}-x_{\mathrm{AI}})}\right )
\right]_{x_{\mathrm{AI}} = x_{\mathrm{AI}_2}}^{x_{\mathrm{AI}} = x_{\mathrm{AI}_1}} \, , \\
\Delta \widetilde{\phi}^\mathrm{LMT}_{\mathrm{grad},\mathcal{E}} (\omega) & =  \frac{1}{2} k_{\rm eff} \, T^2 \,  \omega^2 \,
 \left [ - \frac{1}{2 \mi \omega} \widetilde{h}_{00} (\omega,x_{\mathrm{AI}}) \left (    
 K_+ (\omega)e^{\mi \omega (x_{\mathrm{AI}}-x_{\mathrm{L}_1})} -K_- (\omega)  e^{\mi \omega (x_{\mathrm{L}_2}-x_{\mathrm{AI}})}\right )  
\right]_{x_{\mathrm{AI}} = x_{\mathrm{AI}_2}}^{x_{\mathrm{AI}} = x_{\mathrm{AI}_1}} \, , \\
\Delta \widetilde{\phi}^\mathrm{LMT}_{\mathrm{grad},\mathcal{S}} (\omega) & = \frac{1}{2} k_{\rm eff} \, T^2 \,  \omega^2 \,
 \left [ K_+ (\omega) \Delta \widetilde{\mathcal{T}}^+_{\mathcal{S}} (\omega,x_{\mathrm{L}_1},x_{\mathrm{AI}})  -  K_- (\omega) \Delta \widetilde{\mathcal{T}}^-_{\mathcal{S}} 
(\omega, x_{\mathrm{L}_2},x_{\mathrm{AI}}  )
\right]_{x_{\mathrm{AI}} = x_{\mathrm{AI}_2}}^{x_{\mathrm{AI}} = x_{\mathrm{AI}_1}} \, ,
\end{aligned}
\end{equation}
respectively.\footnote{We work in natural units (i.e. $\hbar = c = 1$) and use the convention $\eta_{\mu\nu} = \mathrm{diag}(-1,1,1,1)$ for the flat spacetime metric.}
Here, $T$ is the interrogation time, $k_\mathrm{eff} = n \omega_a$ is the effective momentum transferred after the beamsplitter sequence, $\hat{\mathbf{n}} = n^i$ is the unit vector defining the orientation of the baseline, $x_\mathrm{AI_1}$ and $x_\mathrm{AI_2}$ are the positions of the AIs along the baseline, and $x_\mathrm{L_1}$ and $x_\mathrm{L_2}$ are the positions of the lasers, with $L = x_\mathrm{L_2}-x_\mathrm{L_1}$ being the length of the baseline. Here we have conveniently chosen a non-inertial frame where $\hat{\mathbf{n}}$ is time-independent. The detector response kernels are defined in terms of experimental parameters as
\begin{equation} \label{eq:KLMTall}
\begin{aligned}
K_{+}(\omega)
& = \frac{e^{\mi\omega T}}{\mathrm{sinc}(\omega L)} \mathrm{sinc}\left(\frac{\omega T}{2}\right) \mathrm{sinc}\left(\frac{\omega \,n L}{2}\right)
\mathrm{sinc} \left (\frac{\omega (T-(n-2) L)}{2} \right )\left(1-\frac{(n-2) L}{T}\right)  \,  ,
\\
K_{-}(\omega) & =  \frac{e^{\mi\omega T}}{\mathrm{sinc}(\omega L)} \mathrm{sinc}\left(\frac{\omega T}{2}\right) \mathrm{sinc}\left(\frac{\omega \,n L}{2}\right) \mathrm{sinc} \left ( \frac{\omega (T-nL)}{2} \right )\left(1-\frac{nL}{T}\right) 
\, ,
\end{aligned}
\end{equation}
where the sign depends on whether the photons are traveling from $x_\mathrm{L_1}$ to $x_\mathrm{L_2}$ ($+$) or vice versa ($-$).\footnote{Here and elsewhere in this paper, $\mathrm{sinc}(x) \equiv \sin (x)/x$.}
While the Einstein phase shift only depends on the {$00$-component} of the metric perturbation, the Doppler phase shift depends on the $\mathcal{O}(h)$ displacement of the atoms initially at a spacetime point $(t,x)$, namely 
\begin{equation}\label{eq:tidal_displacement}
\delta x^{i}(t,x) = - \eta^{ij}\int_{t_0}^t dt' \int_{t_0}^{t'} dt'' \, \left (\partial_0 h_{j0}(t'',x) -\frac{1}{2} \partial_j h_{00}(t'',x) \right ) + \mathcal{O}(hk_\mathrm{eff}/m_a) \, ,
\end{equation}
\end{widetext}
where $t_0$ is the coordinate time at which the sequence is initiated and $m_a$ is the mass of the atom.
The Shapiro phase shift depends on the $\mathcal{O}(h)$ correction to the photon propagation time between emission (i.e. the laser at coordinate position $x_1$) and absorption (i.e. the relevant atomic wavepacket at coordinate position $x_2$), namely
\begin{equation}\label{eq:TScurl}
    \Delta \mathcal{T}_\mathcal{S}^\pm \left (t, x_1, x_2 \right)\equiv \pm\int_{x_1}^{x_2}  \mathcal{H}_{\pm}\left (t\pm(x'-x_1), x'\right )dx' \, ,
\end{equation}
where we defined
\begin{equation}\label{eq:mathcal_H}
    \mathcal{H}_{\pm}\equiv \frac{1}{2}\left(h_{00}\pm 2h_{0i}n^i+h_{ij}n^in^j\right) \, .
\end{equation}

In sections~\ref{sec:ultraheavy}--\ref{sec:ultralight} we will use these master equations to derive the leading-order phase shift caused by DM-induced spacetime perturbations.

\section{Dark Clumps} \label{sec:ultraheavy}

Before proceeding to a detailed signal and sensitivity analysis in sections~\ref{sec:ultraheavy:signal} and \ref{sec:ultraheavy:sensitivity}, respectively, we begin by providing a heuristic estimate of the signal induced by a dark clump. In section~\ref{sec:ultraheavy_heuristics}, we provide a non-relativistic estimate for the induced signal, which can be understood in terms of forces acting on the atoms and is similar to the arguments found in Refs.~\cite{Seto:2004zu,Hall:2016usm, Kawasaki:2018xak,Jaeckel:2020mqa,Baum:2022duc}. As shown in section~\ref{sec:ultraheavy:signal}, this is in agreement with the full general relativistic calculation in Newtonian gauge: the AG observable is dominated by the Doppler time delay, i.e. the time delay due to the tidal displacement of the atoms along the baseline.

\subsection{Heuristics}\label{sec:ultraheavy_heuristics}

A transient clump of non-baryonic matter induces a time-varying acceleration on test masses. Since the atoms in an AG of baseline length $L$ are in free-fall, we may model an AG as an accelerometer. The maximal relative acceleration that can be induced by an effectively point-like clump of DM of mass $M$ is given by 
\begin{equation}\label{eq:rel_acc}
\Delta a \sim \frac{G M}{b^2}\mathrm{min} \left (1, \frac{2L}{b}\right ) \, ,
\end{equation}
where $G$ is Newton's constant and $b$ is the impact parameter (i.e. the minimum distance between the clump and one of the two AIs). 
For a DM clump traveling at speed $v\sim10^{-3}$, as would be expected for virialised DM in our galaxy, the relative acceleration will inherit a non-trivial time dependence. In turn, the FT of the relative acceleration is flat at angular frequencies smaller than $\omega_\mathrm{cut} \sim v/b$ and falls off at higher frequencies. By further invoking signal power conservation, one can estimate the induced relative acceleration as $|\Delta \widetilde{a}| \sim |\Delta a|/\omega_\mathrm{cut}$. Since the AI phase shift induced by forces acting on the atoms is approximately $k_\mathrm{eff}T^2 a$~\cite{Kasevich:1991zz,Storey:1994oka}, where $a$ is the atom's acceleration, the amplitude of the FT of the AG phase shift for frequencies $\omega \lesssim \omega_\mathrm{peak} \sim \mathrm{min}(1/T,\omega_\mathrm{cut})$ can be approximated as
\begin{equation} \label{eq:estdphi}
|\Delta \widetilde{\phi}_\mathrm{grad}| \sim  k_\mathrm{eff} T^2 |\Delta \widetilde{a}| = k_\mathrm{eff} T^2 \frac{G M}{b v}\mathrm{min} \left (1, \frac{2L}{b}\right ) \, .
\end{equation}
Here, $1/T$ is approximately the largest frequency to which a detector is sensitive: indeed, since AGs integrate signals over a time $\sim 2T$, these devices act as a low-pass filter for frequencies $\omega \lesssim 1/T$.
The signal-to-noise ratio for a shot-noise limited device with phase noise power spectral density (PSD) $S_n$ can then be estimated as
\begin{equation}
\begin{aligned}\label{eq:toy_SNR}
\mathrm{SNR} & \sim \sqrt{\frac{|\Delta \widetilde{\phi}_\mathrm{grad}|^2}{S_n} \frac{\Delta \omega}{2\pi}} \\
& = k_\mathrm{eff}T^2\frac{G M}{b v} \sqrt{\frac{1}{S_n}}\sqrt{\frac{\Delta \omega}{2\pi}} \, \mathrm{min} \left (1, \frac{2L}{b}\right ) \, ,
\end{aligned}
\end{equation}
where the bandwidth is given by $\Delta \omega \equiv \omega_\mathrm{peak} - \omega_\mathrm{min} \geq 0$.

It is advantageous to consider the largest impact parameter that would leave a detectable imprint on the experiment's data stream for fixed clump mass, speed $v$ and experimental parameters. For one such event to occur during the measurement campaign, which we assume lasts for a time $T_\mathrm{int}$, i.e. 
\begin{equation}
\frac{\rho_\mathrm{cl}}{M} 
b^2 T_\mathrm{int} v \sim 1 \, ,
\end{equation}
where $\rho_\mathrm{cl}=f_\mathrm{DM}\rho_\mathrm{DM}$ and $f_\mathrm{DM}$ is fraction of DM composed of clumps, one can estimate the maximum impact parameter\footnote{We neglect the effect of the Sun’s and Earth’s gravitational potentials on the local DM speed and energy density. By Liouville’s theorem, such effects induce only $\mathcal{O}(1\%)$ corrections to the local DM density $\rho_\mathrm{DM}$ and sub-percent corrections to the DM mean velocity distribution~\cite{Griest:1987vc}. Consequently, the impact parameter is modified at the $\mathcal{O}(1\%)$ level, which is negligible for our purposes.
} as 
\begin{equation}\label{eq:toy_bmax}
b \sim \sqrt{\frac{M}{\rho_\mathrm{DM}f_\mathrm{DM}} \frac{1}{v T_\mathrm{int}}} \, .
\end{equation} 
Using Eqs.~\eqref{eq:toy_SNR} and \eqref{eq:toy_bmax} and solving for $f_\mathrm{DM}$, one can then estimate the minimum fraction of dark matter composed of dark clumps of mass $M$. Crucially, the smallest dark matter fraction that can be probed by an AG scales with the clump mass as
\begin{equation}\label{eq:toy_scalings}
f_\mathrm{DM}^{(\mathrm{min})} \propto 
\setlength{\arraycolsep}{6pt}
  \renewcommand{\arraystretch}{1.2}
  \left\{\begin{array}{l @{\quad} l r l}
        M^{-1} & \text{for $b 
        \lesssim v T \, , \, L $ } & \text{(A)} \\
        M^{-1/3} & \text{for $v T \lesssim b
        \lesssim L $} & \text{(B)} \\
        M^0 & \text{for $ L \lesssim b 
        \lesssim v T$} & \text{(C)} \\
        M^{1/5} & \text{for $ b
        \gtrsim v T \, , \, L $} & \text{(D)} \\
        M & \text{for $b
        \sim v/\omega_{\mathrm{min}}$} & \text{(E)}
  \end{array}\right.
\end{equation}
where condition (E) is derived from the requirement that the signal be contained in the experimental bandwidth. Note that regimes (A) and (E) can also be observed in the sensitivity curves of PTAs~\cite{Dror:2019twh}. For a more detailed comparison, see the discussion at the end of section~\ref{sec:ultraheavy:sensitivity}.

From Eq.~\eqref{eq:toy_scalings} and Table~\ref{tab:exps}, one can straightforwardly estimate the scaling behaviour of the projected reach for specific experiments. The projected sensitivity curve of any terrestrial experiment (i.e. $L \ll v T$) is then characterised by a clump mass dependence exhibited by regimes (A), (C), (D) and (E). This is to be contrasted with space-based experiments (i.e. $L \approx v T$), whose sensitivity curves display the scaling behavior of regimes (A), (D) and (E). For both types of detector the maximum reach is therefore expected to occur at $b 
\sim L$. From Eq.~\eqref{eq:toy_scalings}, one can also gain insight into the qualitative behavior of $f_\mathrm{DM}^{(\mathrm{min})}$ in the limits $M\rightarrow 0$ and $M\rightarrow \infty$. In the small mass limit, a clump leaves a detectable signal if its maximum impact parameter is small, which in turn implies that dark clumps would constitute a large DM overdensity. In the large-mass limit, instead, a heavy clump leaves a detectable signal in the detector for much larger impact parameters; however, encounters with heavier clumps are less likely to occur during an experiment's run, unless a large DM overdensity is populated by such bound states. 

Furthermore, one can also easily estimate the peak sensitivity of proposed experiments. Using Eqs.~\eqref{eq:toy_SNR}-\eqref{eq:toy_bmax} and assuming a $\mathrm{SNR} \sim 1$, the peak sensitivity occurs at clump masses
\begin{equation}\label{eq:estimate_M}
\begin{aligned}
M^{(\star)} \sim \, & \mathcal{O}(10^6 \, \mathrm{kg})
\left( \frac{150 \, {\rm s}}{T} \right)^{2}  \left( \frac{L}{4.4 \times 10^7 \, {\rm m}} \right)^{3/2} \\
& \quad \times \left( \frac{250}{n} \right) \left( \frac{S_n}{10^{-8} \, {\rm Hz}^{-1}} \right)^{1/2}\, ,
\end{aligned}
\end{equation}
and dark matter fraction
\begin{equation}
\begin{aligned}\label{eq:estimate_fDM}
f_\mathrm{DM}^{(\star)} \sim\, & \mathcal{O}(10^{-1})
\left( \frac{150\, {\rm s}}{T} \right)^{2}\left( \frac{250}{n} \right) \left( \frac{10^8 \, \mathrm{s}}{T_{\rm int}} \right)\,\\
& \times
\left( \frac{4.4 \times 10^7 \,{\rm m}}{L} \right)^{1/2} \left( \frac{S_n}{10^{-8} \, {\rm Hz}^{-1}} \right)^{1/2}\, ,
\end{aligned}
\end{equation}
where $n$ is the number of LMT kicks.
Interestingly, the peak occurs at clump masses that lie outside of the mass range accessible to complementary laboratory probes (e.g., laser interferometers and optomechanical sensors).
In light of the promising reach and unique mass range, in the following subsections we calculate the full AG phase shift signal using the formalism introduced in section~\ref{subsec:formalism}. As we show in the following subsections, the estimate provided here is in excellent agreement with the full calculation and can be used to understand the qualitative features of the reach plot (cf.~section~\ref{sec:ultraheavy:sensitivity}). 

\subsection{Full signal calculation}\label{sec:ultraheavy:signal}

In this subsection, we compute the full general relativistic contributions to the AG observable to leading order in the metric perturbation and the atom recoil velocity. As we explicitly show, the signal is dominated by the Doppler gradiometer phase shift, which accounts for the time delay due to the tidal displacement of the atoms along the baseline.

We begin the general relativistic calculation by computing the metric perturbation sourced by a transient DM clump. Assuming that the clump radius is much smaller than all the relevant length scales in the problem, we model the bound state as a point-like particle. In an inertial frame where the clump is moving at non-relativistic speeds, the leading-order non-zero entries of $h_{\mu\nu}$ read as
\begin{equation}\label{eq:metric_dynamic}
\begin{aligned}
    h_{00} = -2\Phi \, , \, h_{ij} = -2\Phi \delta_{ij} \, , h_{0i} = - 4 {v_i} \Phi 
    \, ,
\end{aligned}
\end{equation}
where $\Phi$ is the weak Newtonian potential sourced by the body (with a mass $M$) and $v_i = \mathbf{v}$ is the clump velocity (see Appendix D of Ref.~\cite{Badurina:2024rpp} for a derivation). For a clump moving with impact parameter $\vec{b}$ relative to the origin, the value of the potential at $x^\mu = (t,\vec{x})$ can be expressed as
\begin{equation}\label{eq:NewPot_x}
\begin{aligned}
\Phi(t, \vec{x}) & = -\frac{G M}{|\vec{b}+\vec{v}(t-t_s)-\vec{x}|} \\
& = -\frac{G M}{\sqrt{r_\perp^2(\vec{x}) + r_\parallel^2(t,\vec{x}) }}  \, ,
\end{aligned}
\end{equation}
where $t_s$ is the time at which the body is closest to the origin. In this expression, $G$ is Newton's constant, $\vec{r}(t,\vec{x})=\vec{b}+\vec{v}(t-t_s)-\vec{x}$, and $r_\perp$ and $r_\parallel$ are the magnitude of the components of $\vec{r}$ perpendicular and parallel to $\vec{v}$, respectively.\footnote{By definition, $\vec{b}\cdot \vec{v} = 0$.} Explicitly, these components may be expressed as
\begin{equation}
\begin{aligned}
\vec{r}_\perp(\vec{x}) & =  \vec{b}+(\hat{\vec{v}} \cdot \vec{x}) \, \hat{\vec{v}} - \vec{x} \, , \\
\vec{r}_\parallel(t,\vec{x}) & = \vec{v} (t-t_s) - (\hat{\vec{v}}_s \cdot \vec{x}) \, \hat{\vec{v}} \, ,
\end{aligned}
\end{equation}
from which one can immediately note that only the component of $\vec{r}$ parallel to the clump's velocity is time-dependent.
Using~Equations~\eqref{eq:total} and \eqref{eq:LMT_tot_F} and working in the limit in which the distance from the AI to the nearest laser is negligible with respect to the length of the baseline (i.e. setting $x_\mathrm{L_1} = x_\mathrm{AI_1}$ and $x_\mathrm{L_2} = x_\mathrm{AI_2}$ in Eq.~\eqref{eq:LMT_tot_F}), the three contributions to the total atom gradiometer phase shift read as
\begin{widetext}
\begin{equation} \label{eq:point_source_phase}
\begin{aligned}
\Delta \widetilde{\phi}^\mathrm{LMT}_\mathrm{grad,\mathcal{D}}(\omega) &=
\frac{1}{2}\, k_{\rm eff} \, T^2 \, \hat{\vec{n}} \cdot \left( \nabla - \, 4 \, \mi \omega \, \Vec{v} \right) \,
\Big[ \big( K_+ + e^{\mi\omega L} K_- \big) \widetilde{\Phi}(\omega, \Vec{x}_{\mathrm{AI}_1})
- \big( K_+ + e^{-\mi\omega L} K_- \big) e^{\mi \omega L} \widetilde{\Phi}(\omega, \Vec{x}_{\mathrm{AI}_2})
 \Big]   \, , \\
 \Delta \widetilde{\phi}^\mathrm{LMT}_\mathrm{grad,\mathcal{E}}(\omega) & = 
- \frac{\mi }{2} \, k_{\rm eff} \, T^2 \, \omega \,
\Big[ \big( K_+ - e^{\mi\omega L}K_- \big) \widetilde{\Phi}(\omega, \Vec{x}_{\mathrm{AI}_1})
 - \big( K_+ - e^{- \mi\omega L} K_- \big) e^{\mi \omega L} \widetilde{\Phi}(\omega, \Vec{x}_{\mathrm{AI}_2})
 \Big] \, , \\ 
\Delta \widetilde{\phi}^\mathrm{LMT}_\mathrm{grad,\mathcal{S}}(\omega) &= 
 - \frac{1}{2}\,  k_{\rm eff} \, T^2
 \, \omega^2 \, 
 \Big[ K_{+} \Delta \widetilde{\mathcal{T}}_\mathcal{S}^+ (\omega, \Vec{x}_{\mathrm{AI}_1}, \Vec{x}_{\mathrm{AI}_2})
 + K_- \Delta \widetilde{\mathcal{T}}_\mathcal{S}^- (\omega, \Vec{x}_{\mathrm{AI}_2}, \Vec{x}_{\mathrm{AI}_1})
 \Big] \, ,
\end{aligned}
\end{equation}
\end{widetext}
where for brevity we omitted the explicit $\omega$-dependence of the response kernels and defined $\nabla \widetilde{\Phi}(\omega, \vec{x}_\mathrm{AI_1}) \equiv \nabla \widetilde{\Phi}(\omega, \vec{x}) |_{\vec{x}_\mathrm{AI_1}}$.
The phase shift takes inputs from the FT of the Newtonian potential defined in Eq.~\eqref{eq:NewPot_x}, namely 
\begin{equation}\label{eqn:potentialFT}
\widetilde{\Phi} ( \omega,\mathbf{x}) =  -\frac{2GM}{v}\, 
e^{-\mi \omega \left (t_s + \frac{\hat{\vec{v}} \cdot \mathbf{x}}{v}\right )} K_0 \left( \frac{r_{\perp}(\vec{x})}{v}\, \omega \right) \, ,
\end{equation} 
and its spatial gradient, 
\begin{widetext}
\begin{equation} \label{eqn:gradpotentialFT}
\begin{split}
& \nabla \widetilde{\Phi}  ( \omega, \mathbf{x})  =   -\frac{2GM}{v^2}\, \omega \,
e^{-\mi \omega \left (t_s + \frac{\hat{\vec{v}} \cdot \mathbf{x}}{v}\right )} \Bigg[ 
 \hat{\vec{r}}_{\perp}(\vec{x}) \, K_1 \left( \frac{r_{\perp}(\vec{x})}{v}\, \omega \right) - \mi\, \hat{\vec{v}}\,K_0 \left( \frac{r_{\perp}(\vec{x})}{v}\, \omega \right) \Bigg] \, ,
\end{split}
\end{equation} 
where $K_0(x)$ and $K_1(x)$ are modified Bessel functions of the second kind. 
Moreover, the Shapiro gradiometer phase shift depends on the Shapiro time delays, which read as
\begin{equation}\label{eq:Newt_Shapiro}
\begin{aligned}  
       \Delta \mathcal{T}_\mathcal{S}^+(t,  \Vec{x}_{\mathrm{AI}_1}, \Vec{x}_{\mathrm{AI}_2})
       = -2 (1+ 2\, \vec{v}\cdot\hat{\vec{n}}) \int_{-L/2}^{L/2}\Phi\left(t+\frac{L}{2}+ x',\Vec{x}_\mathrm{mid}+ x'\hat{\vec{n}}\right) dx' \, , \\
       \Delta \mathcal{T}_\mathcal{S}^-(t, \Vec{x}_{\mathrm{AI}_2}, \Vec{x}_{\mathrm{AI}_1})
       = -2 (1- 2\, \vec{v}\cdot\hat{\vec{n}}) \int_{-L/2}^{L/2}\Phi\left(t+\frac{L}{2}- x',\Vec{x}_\mathrm{mid}+ x'\hat{\vec{n}}\right) dx'\, ,
\end{aligned}
\end{equation}
where $\Vec{x}_\mathrm{mid} \equiv (\Vec{x}_{\mathrm{AI}_1} + \Vec{x}_{\mathrm{AI}_2})/2 $ is the midpoint of the baseline. To make progress, let us introduce the vector $\mathbf{n}'_\pm \equiv \hat{\vec{n}}\mp \vec{v} $ with $n'_\pm \equiv | \mathbf{n}'_\pm | \neq 1$, such that the Newtonian potential in Eq.~\eqref{eq:Newt_Shapiro} may be expressed as
\begin{equation}
\Phi\left(t+\frac{L}{2}\pm x',\Vec{x}_\mathrm{mid}+ x'\hat{\vec{n}}\right) = - \frac{G M}{n_\pm'\sqrt{(r'_\parallel - x')^2+(r'_\perp)^2}} \, ,
\end{equation}
where $\vec{r}' = (\vec{b} + \vec{v}(t-t_s+L/2)-\vec{x}_\mathrm{mid})/n_\pm'$, and $\vec{r}'_\parallel = (\vec{r}'\cdot \hat{\vec{n}}'_\pm) \hat{\vec{n}}'_\pm$ and $\vec{r}'_\perp = \vec{r} - \vec{r}'_\parallel$ are the components of $\vec{r}'$ parallel and perpendicular to $\vec{n}'_\pm$. Following the steps in Refs.~\cite{Siegel:2007fz,Du:2023dhk} and neglecting the (parametrically suppressed) contribution proportional to $\pm \mathbf{v}\cdot\hat{\mathbf{n}}$, the analytical form of the Fourier transform of the Shapiro time delays may be compactly expressed in two limits: as
\begin{equation}
\begin{split}\label{eq:ShapiroFT_longb}
\Delta \widetilde{\mathcal{T}}_\mathcal{S}^\pm  
&= 4\frac{GML}{v} e^{-\mi\omega ( t_s + \vec{x}_\mathrm{mid} \cdot \hat{\vec{v}}/v-L/2)} 
K_0 \left( \frac{r_{\perp}(\vec{x}_\mathrm{mid})}{v}\, \omega \right)  ,\, \, \text{for~} b \gg L \, ,
\end{split}
\end{equation} 
and as
\begin{equation}
\begin{split}\label{eq:ShapiroFT_shortb}
\Delta \widetilde{\mathcal{T}}_\mathcal{S}^\pm  
&= 4 \frac{ G M \pi}{ \omega \, n_\pm'} e^{-\mi\omega ( t_s -L/2)} 
e^{\mi \omega (\vec{b}^\pm_\perp-\vec{x}^\pm_{\mathrm{mid},\perp})\cdot  \vec{\hat{v}}^\pm_{\perp}/v^\pm_{\perp}} 
e^{- \omega |(\vec{b}^\pm_\perp-\vec{x}^\pm_{\mathrm{mid},\perp})\times \vec{\hat{v}^\pm}_\perp| /v^\pm_{\perp}}  ,\, \,  \text{for~} b \ll L \, ,
\end{split}
\end{equation}
where $\vec{b}^\pm_\perp$, $ \vec{x}^\pm_{\mathrm{mid},\perp}$ and $v^\pm_{\perp}$ are defined using the rule $\vec{A}_\perp^\pm = \vec{A}-(\vec{A}\cdot \hat{\vec{n}}'_\pm)\hat{\vec{n}}'_\pm$.
\end{widetext}

\subsubsection{Frequency-dependence}

Because of the nontrivial frequency-dependence of the the three AG phase shift contributions, we now study the behavior of the signal at low and high frequencies, and the hierarchy between the Doppler, Einstein and Shapiro phase shifts. As anticipated at the beginning of this subsection, we show that the signal is: ($i$) power law and exponentially suppressed for $\omega \gtrsim 1/T$ and $\omega \gtrsim v/b$, respectively, and ($ii$) dominated by the Doppler phase shift, whose spectrum is flat for $\omega \ll v/b$.

Using Eqs.~\eqref{eqn:potentialFT}-\eqref{eq:ShapiroFT_shortb}, one can infer the parametric scaling of the signal induced by a transient clump. Firstly, we shall focus on the high-frequency regime. Since $K_{0,1}(\omega \tau) \propto e^{-\omega \tau}/\sqrt{\omega}$ for $\omega \tau \gtrsim 1$, $\widetilde{\Phi}$, $\nabla \widetilde{\Phi}$ and $\Delta \mathcal{T}_\mathcal{S}^\pm$ are all exponentially suppressed at frequencies $\omega \gtrsim 1/\tau \sim \min ( v/b,v/L)$.\footnote{More precisely: $\widetilde{\Phi}(\omega \tau)$ and $\Delta \mathcal{T}_\mathcal{S}^\pm(\omega \tau)$, the latter in the limit $b\gg L$, are proportional to $e^{-\omega \tau}/\sqrt{\omega}$; $\nabla \widetilde{\Phi}(\omega \tau) \propto e^{-\omega \tau} \sqrt{\omega}$ and, in the limit $b\ll L$, $\Delta \mathcal{T}_\mathcal{S}^\pm(\omega \tau) \propto e^{-\omega \tau} / \omega$.} As anticipated in the heuristic estimate in section~\ref{sec:ultraheavy_heuristics}, this follows from the transient nature of the signal: from simple dimensional analysis, the only timescales that can appear in the problem depend on the velocity of the clump, and on the distance scales $b$ and $L$. Furthermore, for $\omega T, \omega nL \gg 1$, the detector response kernels (cf. Eq.~\eqref{eq:KLMTall}) scale as $K_\pm \propto n^{-1}/(\omega T)^2$. This power-law suppression is a direct consequence of the fact that an AG integrates a signal on a timescale $\sim T$, making an AG a low-pass filter for signals oscillating at $\omega \lesssim 1/T$. Consequently, at sufficiently high frequency, the modulus of all contributions to the gradiometer phase shift is exponentially and power law suppressed. 

At low frequencies, the hierarchy between the different contributions to the gradiometer phase shift is manifest. Since $\omega K_1(\omega \tau) \approx 1/\tau$ and $\omega^n K_0(\omega \tau) \approx - \omega^n \ln (\omega \tau)$ for $\omega \tau \ll 1$ and $n>0$, the FT of the spatial gradient of the potential sourced by the clump is dominated by $K_1$ and is therefore constant. For example, setting the origin of the coordinate system at one of the AIs, $|\nabla \widetilde{\Phi}| \sim GM/v b$ for $\omega \ll v/b$, which agrees with our earlier estimate. Furthermore, in the limit $\omega n L, \omega T \ll 1$, the gradiometer phase shift contributions depend on two combinations of the detector response functions, namely $K_+ - K_- \sim 2L/T \ll 1$ and $K_+ + K_- \sim 2(1-(n-1)L/T)) \approx 2$ for all experimental configurations considered in Table~\ref{tab:exps}. The former appears in the Doppler and Shapiro gradiometer phase shifts, when neglecting the subdominant $(\vec{v}\cdot\hat{\vec{n}})$-dependent term in the Shapiro time delay, while the latter only appears in the Einstein term. Notably, this implies that, in the limit $b\gg L$, the modulus of the Einstein and Shapiro phase shifts are parametrically suppressed by $\sim \omega L^2 /T $ and $\sim \omega^2 L b |\ln(\omega b(1+L/b)/v)|$, respectively. In the limit $b\ll L$, instead, these contributions are parametrically suppressed by $\sim \omega  b L |\ln(b/L)| /T $ and $\sim \omega v b$, respectively. 

In light of the $\omega$-dependence of the phase shift, we now investigate the maximum contribution of the Einstein and Shapiro phase shifts to the total gradiometer observable. Note that the global maximum of $\omega^n K_0(\omega \tau)$ and $\omega e^{-\omega\tau}$ occurs at $\omega \sim 1/\tau$. Hence, we shall focus on $\omega \sim v/b$. In the regime $ nL, T \ll b/v$ and relative to the Doppler contribution, the gradiometer phase shift contributions depend on the two combinations of the detector response functions previously discussed. Hence, the Einstein term is always suppressed by $\sim v L/T$, while the Shapiro term is suppressed by $\sim v^2 \min (1,L/b)$. In the regime $ nL, T \gg b/v$, however, the detector response kernels scale as $1/(\omega T)^2$; hence, the suppression of the Shapiro term is unchanged, while the Einstein term is now only $v$-suppressed relative to the Doppler contribution. Since $nL \ll T$ in all experiments considered here (cf. Table~\ref{tab:exps}), we now consider the regime $nL \ll b/v \ll T$. In this case, the detector response kernels are also power-law suppressed, and the Einstein and Shapiro contributions are similarly suppressed relative to the Doppler part. 

Since the Einstein and Shapiro contributions are parametrically suppressed in all of the regimes of interest, we conclude that the gradiometer observable will be dominated by the Doppler term, i.e. by the phase shift associated with the tidal displacement of the atoms along the baseline.

\subsection{Sensitivity analysis \& projected reach}\label{sec:ultraheavy:sensitivity}

As shown in the previous subsection, a nearby and massive transient dark clump induces a time-dependent phase shift, which can therefore be analyzed in the frequency domain. To extract small signals from the background, we use the matched filtering prescription~\cite{Moore:2014lga}, a procedure that is commonly employed by the gravitational wave community to analyze transient gravitational wave signals. Assuming that the time interval between successive measurements is much shorter than the duration of the signal, which we take to be much shorter than the total measurement campaign,\footnote{These conditions allow us to neglect finite-sampling and finite-time effects. If the time interval between successive measurements satisfies $\Delta t \ll b/v$, the spectral content of the signal is not affected by spectral distortions due to aliasing. If the duration of the total measurement campaign satisfies $T_\mathrm{int} \gg b/v$, the signal does not peak in the zeroth frequency bin, which is degenerate with static effects. Although the latter is easily satisfied, the former criterion requires interleaved interferometer sequences~\cite{PhysRevLett.81.5780,Savoie:2018guq}.} the square of the signal-to-noise (SNR) ratio is given by
\begin{equation}\label{eqn:SNR}
\mathrm{SNR}^2 = 4 \int_{\omega_{\rm min}}^{\omega_{\rm max}} \frac{d\omega}{2\pi} \frac{\left |\Delta \widetilde{\phi}_\mathrm{grad}^\mathrm{LMT} (\omega) \right |^2}{S_{n} (\omega)},
\end{equation}
where $\Delta \widetilde{\phi}(\omega)_\mathrm{grad}^\mathrm{LMT}$ is the total gradiometer observable (cf. Eq.~\eqref{eq:LMT_tot_F}) and $S_n$ is the power spectral density (PSD) of the gradiometer phase noise; here, we assume that the gradiometer phase noise is shot-noise limited between $\omega_{\rm min}$ and $\omega_{\rm max}$ (cf. Table~\ref{tab:exps}). Since shot noise is Gaussian, the probability density function of the $\mathrm{SNR}^2$ in the absence of a signal is given by a chi-squared distribution with one degree of freedom~\cite{Maggiore:2007ulw}. Setting the probability of false alarm at $5\%$, the threshold value of the $\mathrm{SNR}^2$, which we define as $\mathrm{SNR}^2_t$, is approximately 4. 

To estimate the projected reach, we apply a statistical procedure previously presented in the context of pulsar timing arrays~\cite{Dror:2019twh, Ramani:2020hdo, Lee:2020wfn} and laser interferometer searches of DM clumps~\cite{Du:2023dhk}. We perform Monte Carlo simulations to sample the DM initial conditions, namely the DM clump position and velocity vector.\footnote{For related works on Monte Carlo-based sensitivity studies for DM clump detection using interferometers, see Refs.~\cite{Hall:2016usm,Baum:2022duc}.} We describe the distribution of DM clump velocities using the Standard Halo model (SHM)~\cite{Lewin:1995rx}. This model yields a Maxwell-Boltzmann distribution whose velocity dispersion is set, at the solar position, by the value of the local standard of rest $v_0 \approx 238$~km/s; furthermore, the distribution is boosted by the average speed of the Earth relative to the halo rest frame $v_\mathrm{obs} \approx 252$~km/s, and truncated at the escape velocity $v_{\mathrm{esc}}=600$~km/s~\cite{Baxter:2021pqo}. The spatial clump distribution is assumed to be homogeneous and isotropic. For each choice of the dark matter fraction and clump mass, we generate 200 simulations in a volume centered around the baseline. Since events with minimal impact parameters induce maximal phase shifts (cf. section~\ref{sec:ultraheavy:signal}), in each simulation we compute the phase shift induced by the nearest encounter. From the distribution of minimum impact parameters for a given choice of $M$ and $f_\mathrm{DM}$, we compute the 90\textsuperscript{th} percentile, which we define as $b_\mathrm{min}$. The 90\% upper limit on the dark matter fraction for a given clump mass is then determined by computing the smallest dark matter fraction for which the gradiometer phase shift at $b_\mathrm{min}$ exceeds the signal-to-noise ratio threshold $\mathrm{SNR}_t$. For a detailed discussion on the analytical derivation of the minimum impact parameter, see Appendix~\ref{app:min_impact_parameter}. 

\begin{figure}[t!]
	\includegraphics[width=0.48 \textwidth]{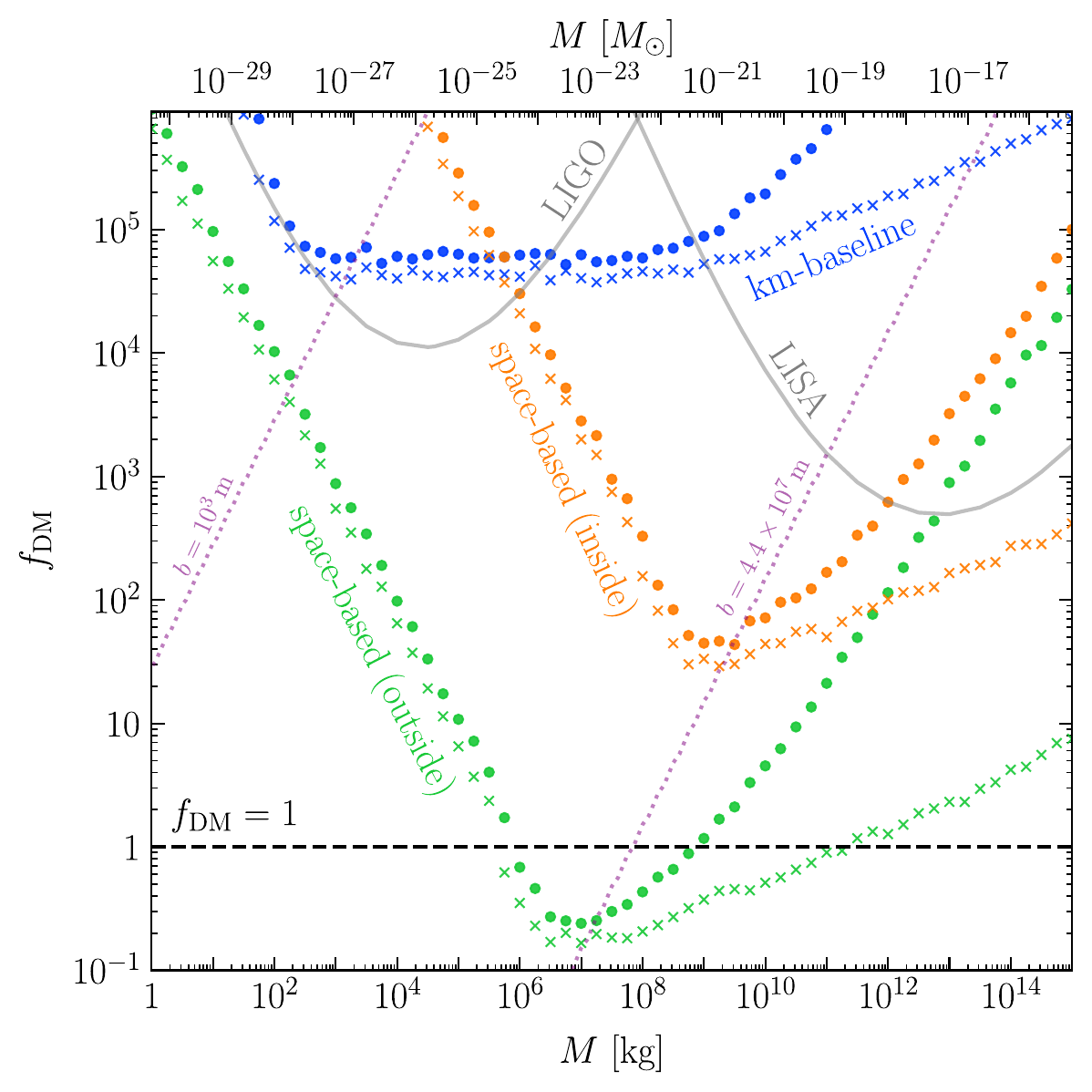} 
    \centering
    \caption{MC-generated projected 90\% upper limits on the fraction of DM ($f_\mathrm{DM}$) composed of transient clumps of mass $M$. We assume an experimental observation time $T_\mathrm{int}=10^8$~s, $\rho_{\mathrm{DM}}=0.46$~GeV/cm$^3$ and a signal-to-noise ratio threshold $\mathrm{SNR}_t \approx 2$. We consider three (broadband) AG concepts: a terrestrial experiment with a km-long baseline (blue), a space-based experiment consisting of two AIs that are confined within two satellites in mid-Earth orbit (orange), and a space-based gradiometer that enables the free evolution of the atoms outside each satellite (green). The parameters for these experiments are summarized in Table~\ref{tab:exps}. For AG experiments, we present projections for both optimistic (marked by crosses) and benchmark (marked by dots) scenarios. Projections for laser interferometer experiments, including LIGO and LISA, as derived in Ref.~\cite{Du:2023dhk}, are also plotted (grey) for comparison. We also plot regions of parameter space where $b$, as estimated in Eq.~\eqref{eq:toy_bmax} and carefully computed in Eq.~\eqref{eqn:b_min}, is equal to $L$.}
    \label{fig:UHDMreach}
\end{figure}

In Figure~\ref{fig:UHDMreach}, we plot the 90\% upper limit on the dark matter fraction saturated by clumps as a function of the clump mass, focusing on the projected reach of the broadband AG experiments described in Table~\ref{tab:exps}. For each experiment we plot the projected reach assuming a benchmark and optimistic experimental bandwidth, which we display with dots and crosses, respectively. Since the Doppler phase shift is the dominant contribution to the gradiometer observable in the entirety of parameter space, we find that the qualitative scaling of the projected sensitivity curves, which are based on the full calculation presented in section~\ref{sec:ultraheavy:signal}, agrees with the heuristic estimate from section~\ref{sec:ultraheavy_heuristics}. For example, we find that the projected reach of space-based proposals peaks at $b_\mathrm{min}
\sim L$, while the reach of the Earth-based experiment features a flat spectrum, since $L\ll v T$, where $v\sim v_0 \sim 10^{-3}$. Furthermore, since the leading order gravitational phase shift scales like $k_\mathrm{eff} T^2 a$, where $a$ is the atom acceleration, we find that the space-based proposal enclosing the largest spacetime area (i.e. $\sim k_\mathrm{eff}T^2$) sets the strongest constraints. 
For all AG proposals, note that the optimistic and benchmark curves scale with $M^{1/5}$ and $M$, respectively, when $b\gg L$. This agrees with regimes (D) and (E) in section~\ref{sec:ultraheavy_heuristics}. This should come as no surprise: since benchmark scenarios assume much larger red noise than optimistic scenarios, the peak frequency of the signal $\sim v/b$ in benchmark scenarios becomes comparable to $\omega_\mathrm{min}$ for smaller values of $b$; given the dependence of $b$ on $M$ (cf. Eq.~\eqref{eq:toy_bmax}), the reach transitions from regime (D) to (E) at smaller clump masses.

Although a terrestrial proposal would only probe a signal in the presence of a large DM overdensity, we find that the space-based proposal analogous to AEDGE+ may be able to probe an $\mathcal{O}(10\%)$ subcomponent of DM in the $10^6~\mathrm{kg}\lesssim M \lesssim 10^{10}~\mathrm{kg}$ (or equivalently $10^{-25}~M_\odot\lesssim M \lesssim 10^{-21}~M_\odot$) mass window. 
Remarkably, \textit{there exist no other probes which could constrain a DM subcomponent in this mass window exclusively via gravitational interactions}. In fact, existing and proposed laser interferometers such as LIGO and LISA are only expected to probe DM fractions much greater than unity at lower and higher clump masses~\cite{Kawasaki:2018xak, Du:2023dhk}.\footnote{We note that an alternative approach is to assume $f_{\mathrm{DM}} = 1$ and present the projected event rate as a function of the clump mass, as done in Refs.~\cite{Adams:2004pk,Seto:2004zu,Hall:2016usm,Jaeckel:2020mqa,Lee:2022tsw,Baum:2022duc}. These studies find that the number of detectable events with sufficiently large SNR is less than unity over the anticipated lifetime of laser interferometers such as LIGO and LISA. As a result, these works indirectly conclude that such experiments would only be sensitive to large overdensities of DM clumps.} Arrays of quantum-limited mechanical impulse sensors (e.g., Windchime~\cite{Windchime:2022whs}) would probe clump masses much smaller than 1~kg (or equivalently $10^{-30}~M_\odot$)~\cite{Carney:2019pza}. Proposals based on asteroid-to-asteroid ranging~\cite{Fedderke:2021kuy} would instead probe $\mathcal{O}(10)$ dark matter overdensities in the $10^{13}~\mathrm{kg}\lesssim M \lesssim 10^{15}~\mathrm{kg}$ mass range~\cite{Baum:2022duc}. Pulsar-timing arrays and the Square Kilometer Array would instead probe an $\mathcal{O}(1\%)$ DM subcomponent, albeit at clump masses larger than $10^{20}$~kg (equivalently $10^{-11}~M_\odot$)~\cite{Dror:2019twh}. 

\begin{figure}[t]
	\includegraphics[width=0.48 \textwidth]{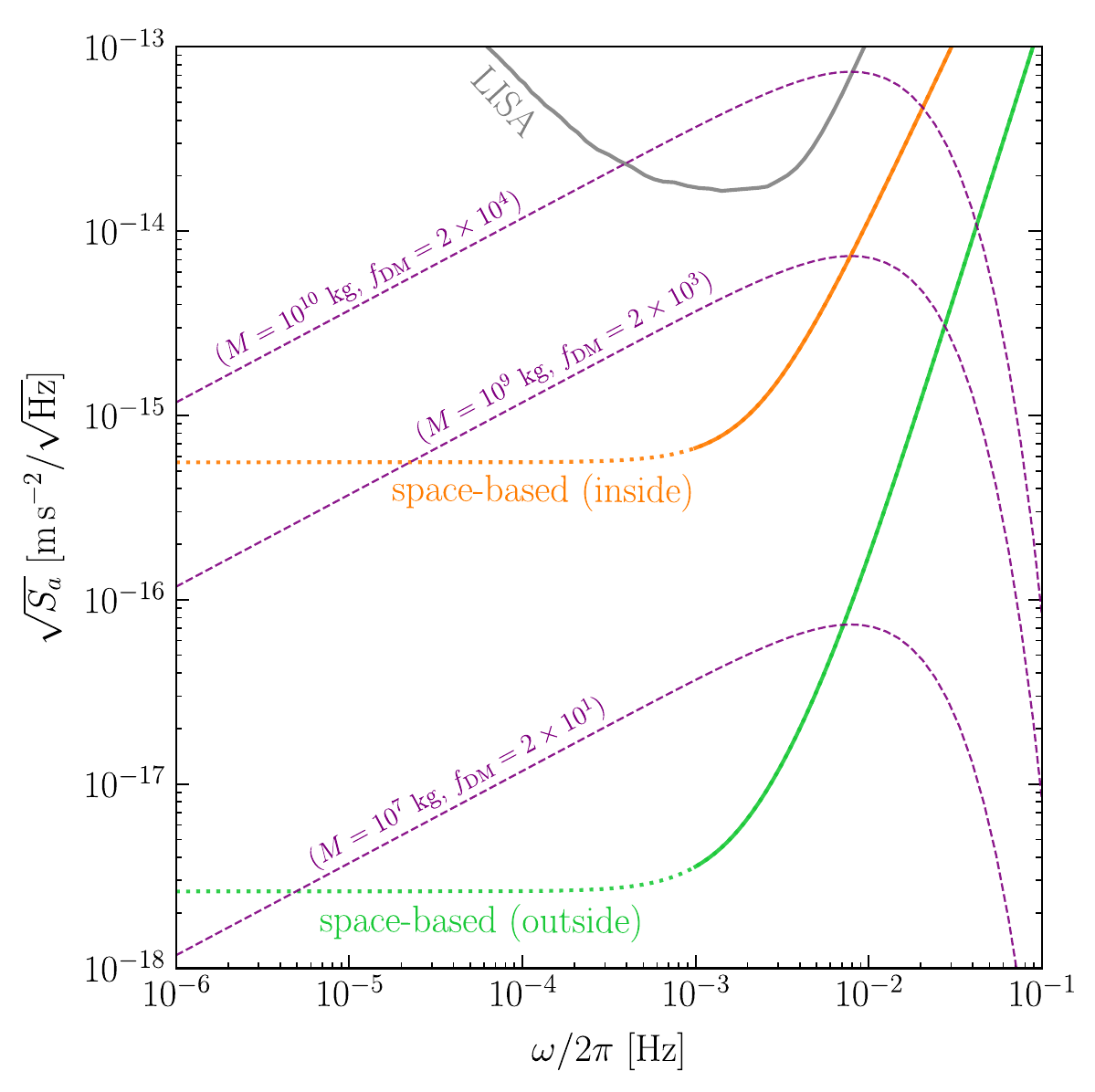} 
    \centering
    \caption{Acceleration sensitivity of LISA (solid gray) and the two space-based AG proposals (green and orange) considered in this work (cf. Table~\ref{tab:exps}) to three events (dashed purple) with identical geometry and impact parameter $b = 4.2\times10^{6}~\mathrm{m}$, but different masses. The solid and dotted lines refer to the benchmark and optimistic sensitivity of proposed AGs. By associating $b$ to the minimum impact parameter defined in Eq.~\eqref{eqn:b_min}, these events correspond to ($M=10^7$~kg, $f_\mathrm{DM}=2\times10^{1}$), ($M=10^9$~kg, $f_\mathrm{DM}=2\times10^{3}$) and ($M=10^{10}$~kg, $f_\mathrm{DM}=2\times10^{4}$) in Fig.~\ref{fig:UHDMreach}.}
    \label{fig:acc_spectrum}
\end{figure}

The impressive reach of proposed space-based AGs stems from their acceleration sensitivity. 
To quantify this, we introduce the square root of the one-sided PSD of the differential acceleration noise, $\sqrt{S_a(\omega)}$, which we define as 
\begin{equation}\label{eq:S_a}
\sqrt{S_a(\omega)} \equiv \frac{\sqrt{S_n}}{\frac{1}{2}k_\mathrm{eff}T^2\left |K_+(\omega)+e^{\mi\omega L}K_-(\omega)\right|}
\end{equation}
for AGs, and as $\sqrt{S_a(\omega)} = L\omega^2 \sqrt{S_h(\omega)}$ for laser interferometers, where $S_h(\omega)$ is the strain noise PSD~\cite{Baum:2022duc}.
In Fig.~\ref{fig:acc_spectrum} we plot the smoothed envelope of $\sqrt{S_a(\omega)}$ using the replacement $\mathrm{sinc}^2 (\omega/\omega^\star) \rightarrow 1/(1+ 2(\omega/\omega^\star)^2)$. For the two space-based AG proposals considered in this work, $\sqrt{S_a} \approx 6\times10^{-16}~\mathrm{m \, s^{-2}}/\sqrt{\mathrm{Hz}}$ and $\sqrt{S_a} \approx 3\times10^{-18}~\mathrm{m \, s^{-2}}/\sqrt{\mathrm{Hz}}$ for $\omega/2\pi \leq 10^{-3}$~Hz. For LISA~\cite{Babak:2021mhe,Robson:2018ifk}, instead, $\sqrt{S_a(\omega)} \approx 2 \times 10^{-14}~\mathrm{m \, s^{-2}}/\sqrt{\mathrm{Hz}}$ for $5\times 10^{-4}~\mathrm{Hz}\lesssim \omega/2\pi \lesssim 3\times 10^{-3}$~Hz.

Parametrically, the frequency dependence and hierarchy of the acceleration sensitivity of space-based AGs and LISA can be understood as follows. For Michelson interferometers, such as LISA, the square root of the strain noise PSD is related to the phase noise PSD $S_{\phi}(\omega)$ by $\sqrt{S_h(\omega)} = \sqrt{S_{\phi}(\omega)}/2 L \omega_{\gamma}$, where $\omega_{\gamma}$ is the laser frequency~\cite{Maggiore:2007ulw}. For $\omega/2\pi \gtrsim 3\times 10^{-3}\,$Hz, LISA is shot-noise limited and $\sqrt{S_{\phi}} = \sqrt{\omega_{\gamma}/\pi P} \approx 3 \times 10^{-4}/\sqrt{\rm Hz}$, where $P$ is the effective laser power \cite{LISA:2017pwj}. Hence, in this frequency range, LISA's acceleration noise spectrum scales with $\omega^2$. Due to instrumental noise, $\sqrt{S_{\phi}} \propto 1/\omega^{2}$ for $5\times 10^{-4}\,$Hz $\lesssim \omega/2\pi \lesssim$ $3\times 10^{-3}\,$Hz, while $\sqrt{S_{\phi}} \propto 1/\omega^3$ at lower frequencies. Consequently, in these frequency bands, LISA's acceleration noise spectrum is flat and scales with $1/\omega$, respectively. This is in contrast to the frequency dependence of the acceleration noise spectrum of AGs. Since, by assumption, the experiments are shot-noise limited, $\sqrt{S_a}$ is flat for $\omega \lesssim 1/T$ and grows as $\omega^2$ at higher frequencies.
To compare the hierarchy of the acceleration sensitivities, we use the ratio of the LISA and AG acceleration noise spectra, which is given by 
\begin{equation}\label{eq:ratio_spectra}
\begin{aligned}
\sqrt{\frac{S_a^{\rm LISA}(\omega)}{S_a^{\rm AG}(\omega)}} & = \sqrt{\frac{S_{\phi}(\omega)}{S_n}}\left (\frac{k_{\rm eff}}{\omega_{\gamma}} \right) \left (\frac{\omega T}{2} \right)^2 \\  & \qquad \times |K_+(\omega) + e^{\mi\omega L} K_-(\omega)| \, .
\end{aligned}
\end{equation}
Since LISA's shot-noise limited phase noise is comparable to that of proposed AGs and $k_{\rm eff}/\omega_{\gamma} \approx n$, in the high-frequency limit, where $(\omega T/2)^2 |K_+ + e^{\mi\omega L} K_-| \to 1/n$, the acceleration sensitivity of space-based AGs becomes asymptotically independent of $n$ and therefore comparable to LISA's. Below $\sim 3\times 10^{-3}\,$Hz, instead, the acceleration sensitivity of space-based AGs is parametrically enhanced by the large number of LMT kicks and small phase noise.

To illustrate how the acceleration sensitivity of an experiment translates into projected bounds on the DM fraction as a function of the clump mass, in Fig.~\ref{fig:acc_spectrum} we plot with dashed curves the square root of the differential acceleration spectrum induced by three sample events. These events share the same geometry\footnote{We assume that the DM clumps travel perpendicularly to the baseline.} and impact parameter, which we set to $ 4.2\times10^{6}~\mathrm{m}$, but differ in clump mass, which we fix to $M=10^7$~kg, $M=10^9$~kg and $M=10^{10}$~kg. By identifying $b$ with the minimum impact parameter (cf.~Eq.~\eqref{eqn:b_min}), these events correspond to points in Fig.~\ref{fig:UHDMreach} where the DM fractions are $f_\mathrm{DM}=2\times10^{1}$, $f_\mathrm{DM}=2\times10^{3}$ and $f_\mathrm{DM}=2\times10^{4}$, respectively. Since $b\ll L$ for both LISA and the space-based AGs, the differential acceleration spectrum induced by such transient clumps applies to each experiment and may be approximated as $|\nabla \widetilde{\Phi}| \sqrt{\omega/2\pi}$, where $\nabla \widetilde{\Phi}$ is defined in Eq.~\eqref{eqn:gradpotentialFT}. Consequently, the detectability of a given event may be inferred by integrating the area between the signal curves and $\sqrt{S_a}$ for each experiment.
From Fig.~\ref{fig:acc_spectrum}, it is then clear that the first event would only be observable by the space-based AG with the best acceleration sensitivity, the second event would also be observable by the less sensitive AG, while the third event would also induce a $\mathrm{SNR} > 1$ in LISA. This agrees exactly with Fig.~\ref{fig:UHDMreach}: the first, second and third events correspond to points in parameter space that lie between the projected upper limits on $f_\mathrm{DM}$ set by the two space-based proposals, between the upper limits set by the ``space-based (outside)" proposal and LISA, and above the LISA curve, respectively.

It is important to note that the promising projections for the ``space-based (outside)'' proposal are based on noise curves that neglect astrophysical foregrounds. As shown in Fig.~\ref{app:fig:noise} (Appendix~\ref{app:noise_sources}), gravitational waves from unresolved galactic and extragalactic binary mergers can significantly degrade the experiment's strain sensitivity. In this case, we expect a substantial reduction in sensitivity across the frequency range $\sim 10^{-5}$--$10^{-2}$~Hz, which corresponds to DM clump signals with impact parameters in the range $5 \times 10^{6}~\mathrm{m} \lesssim b \lesssim 5 \times 10^{9}~\mathrm{m}$. Given this, and in light of Figs.~\ref{fig:UHDMreach} and \ref{app:fig:noise}, we anticipate that an AEDGE$+$-like detector would still achieve stronger (though comparable) constraints on $f_\mathrm{DM}$ than the ``space-based (inside)'' proposal for clump masses between $\sim 10^7$~kg and $\sim 10^{11}$~kg. However, we would still expect this setup to probe $f_\mathrm{DM} \sim 1$ for clump masses around $M \sim 10^6$~kg. Due to uncertainties in the modeling of astrophysical foregrounds, especially in the $10^{-3}$--$10^{-2}$~Hz band (see Appendix~\ref{app:noise_sources}), values of $f_\mathrm{DM} \lesssim 1$ could still be accessible for $M \gtrsim 10^6$~kg.

Finally, we comment on the complementarity of dark clump searches with AGs and PTAs. To facilitate the comparison, we interpret PTAs as laser interferometers: the far mirror is replaced by a pulsar, the baseline length is set by the Earth-pulsar distance, and the photons travel only once along the ``baseline". With this picture in mind, PTAs can be interpreted as pseudo-interferometers with $L\sim$~kpc. PTAs probe frequencies between $\mathcal{O}(10^{-9}~\mathrm{Hz})$ and $\mathcal{O}(10^{-6}~\mathrm{Hz})$, where the range is set by the observation time and the time interval between measurements, respectively. As a result, the PTA observable is most sensitive to events caused by very heavy clumps with impact parameters smaller than the baseline, $b = v/\omega_\mathrm{min} \sim 10^{14}~\mathrm{m}\ll L$. This differs from the AG case, where events with $b$ comparable to the baseline length give rise to the best sensitivity. Consequently, the Shapiro time delay (which, by contrast, dominates the PTA signal in Newtonian gauge at even larger clump masses where $b\ll L$), never exceeds the Doppler contribution in the AG case. Furthermore, when the Doppler time delay dominates the signal in PTAs, the reach scales as $M^{-1}$ and $M$, corresponding to regimes (A) and (E) in Eq.~\eqref{eq:toy_scalings}.

\section{Ultralight Dark Matter}\label{sec:ultralight}

In addition to detecting gravitational signatures from transient DM clumps, atom gradiometers could also measure pressure and energy density fluctuations of ultralight dark matter (ULDM). Before proceeding to a detailed signal and sensitivity analysis in sections~\ref{sec:ultralight:signal} and \ref{sec:ultralight:sensitivity}, respectively, we begin by providing a heuristic estimate of the induced signal.

\subsection{Heuristics}\label{sec:ultralgiht_heuristics}

In light of its large occupation number per de Broglie volume, small mean velocity and velocity dispersion, ultralight dark matter can be described as a temporally and spatially oscillating non-relativistic classical field with frequency largely set by its mass $m$ and amplitude proportional to ${\sim} \sqrt{\rho}/m$, where $\rho$ is the DM energy density. Since gravity couples universally, fluctuations in the energy density and pressure of this field give rise to time-dependent effects on test particles, regardless of non-gravitational interactions between DM and Standard Model fields. These density fluctuations, which arise from the DM's velocity dispersion, can be modeled as quasiparticles with an effective mass $M_\mathrm{eff} \sim \pi^{3/2}\delta \rho \lambda_\mathrm{c}^3$, with $\lambda_\mathrm{c} \sim 1/m v$ the coherence length (or equivalently the de Broglie wavelength)~\cite{Hui:2016ltb}. Neglecting slow-varying contributions, $\delta \rho \sim v^2 \rho \cos(2m t-2k x)$, so that the effective particle's mass oscillates at frequency ${\sim} 2m$. 
From Poisson's equation, the potential sourced by these fluctuations, as experienced by a test mass at a distance $L+\lambda_c$ from the center of the quasiparticle, is given by
\begin{equation}\label{eq:toy_potential}
\begin{aligned}
\Phi(t) & \sim \frac{\pi^{3/2}G M_\mathrm{eff}}{\lambda_c} + \mathcal{O}\left(\frac{L}{\lambda_c}\right) \\
& \sim \frac{\pi^{3/2}G \rho}{m^2}\cos(2mt) + \mathcal{O}\left(\frac{L}{\lambda_c}\right) \, .
\end{aligned}
\end{equation}
Since $h\sim \Phi$ and $\Phi$ is almost monochromatic, one can estimate the signal by replacing the strain induced by a transient GW with $\Phi$ as determined in Eq.~\eqref{eq:toy_potential}. Using the amplitude of the phase shift induced by a transient GW, as derived in Refs.~\cite{Graham:2016plp,Badurina:2024rpp}, we find 
\begin{equation}
\label{eq:toy_ultralight_phase}
\begin{aligned}
\left |\Delta \phi_\mathrm{grad} \right| & \sim \frac{\omega_a}{m^3}\pi^{3/2} G \rho  \\ & \qquad \qquad  \times \min \left ( 1, m^2 T^2 \right ) \min \left (1, m nL \right ) \, .
\end{aligned}
\end{equation}
Note that the ULDM is phase coherent within a coherence time $\tau_c \sim 1/m v^2$, which in turn implies that the signal will have non-zero support in a band $\Delta \omega \sim 2\pi/\tau_c$. In what follows we assume that the experiment is shot-noise limited. If $T_\mathrm{int} \lesssim \tau_c$, the experiment cannot resolve the full spectral features of the signal, and the signal is contained within a frequency bin of size $\Delta \omega \sim 2\pi/T_\mathrm{int}$ and, because of power conservation, $|\Delta \widetilde{\phi}_\mathrm{grad}|^2 \sim |\Delta \phi_\mathrm{grad}|^2 T_\mathrm{int}^2$. If $T_\mathrm{int} \gtrsim \tau_c$, the experiment can resolve the full frequency content of the signal. In this case, the SNR can be estimated via  Bartlett's method~\cite{VanderPlas_2018,Badurina:2021lwr}. This procedure consists of cutting the data stream into segments of duration $\tau_c$ and averaging the PSDs for each chunk. In this case, the signal is contained within a single frequency bin of size $\Delta \omega \sim 2\pi/\tau_c$ and the noise PSD is suppressed by a factor of $ \sqrt{T_\mathrm{int}/\tau_c}$. Because of power conservation, $|\Delta \widetilde{\phi}_\mathrm{grad}|^2 \sim |\Delta \phi_\mathrm{grad}|^2 \tau_c^2$. Therefore, the SNR can be estimated as
\begin{equation}\label{eq:toy_ultralight_SNR}
\begin{aligned}
\mathrm{SNR} & \sim \sqrt{\frac{|\Delta \widetilde{\phi}_\mathrm{grad}|^2}{S_n}\frac{\Delta \omega}{2\pi}} \\
& = \frac{|\Delta \phi_\mathrm{grad}|}{\sqrt{S_n}} 
\begin{cases}
    T_\mathrm{int}^{1/2} & \text{for $T_\mathrm{int} \lesssim \tau_c$} \\
    (T_\mathrm{int}\tau_c)^{1/4} & \text{for $T_\mathrm{int} \gtrsim \tau_c$} \, .
\end{cases}
\end{aligned}
\end{equation}
Setting $\rho = f_\mathrm{DM} \rho_\mathrm{DM}$ and using Eqs.~\eqref{eq:toy_ultralight_phase}-\eqref{eq:toy_ultralight_SNR}, the reach of an AG scales with the DM mass in the limit $T_\mathrm{int}\lesssim \tau_c$ as 
\begin{equation}\label{eq:toy_scalings_ultralgiht}
f_\mathrm{DM}^{(\mathrm{min})} \propto 
\setlength{\arraycolsep}{6pt}
  \renewcommand{\arraystretch}{1}
  \left\{\begin{array}{l @{\, \,} l r l}
        m^{0} & \text{for $m \lesssim 1/T \, ,\, 1/nL $} 
        & \text{(A)} \\
        m & \text{for $  1/nL \lesssim  m \lesssim 1/T$} & \text{(B)} \\ 
        m^2 & \text{for $1/T \lesssim  m \lesssim 1/nL$} & \text{(C)} \\
        m^3 & \text{for $ m \gtrsim 1/T \, , \, 1/nL$} & \text{(D)}
  \end{array}\right.
\end{equation}
where these relations only hold for $\omega_\mathrm{min} \lesssim m \lesssim \omega_\mathrm{max}$ and $L\ll\lambda_c$. In the limit $T_\mathrm{int} \gtrsim \tau_c$, the RHS of each regime is multiplied by $m^{1/4}$. Note that regime (D) also features in PTA searches~\cite{Khmelnitsky:2013lxt}; we elaborate further on the similarities and differences in section~\ref{sec:ultralight:sensitivity}.

From Eq.~\eqref{eq:toy_scalings_ultralgiht} and Table~\ref{tab:exps}, one can straightforwardly estimate the scaling behavior of the reach for specific experiments. 
Since $T > nL$ in all proposals considered in this work, the sensitivity curves will display the scaling behavior of regimes (A) and (C). Regime (A) gives the best reach, and regime (D) would only be displayed by space-based experiments.  Importantly, from Eq.~\eqref{eq:toy_scalings_ultralgiht}, we can also gain insight into the qualitative behavior of $f_\mathrm{DM}^{(\mathrm{min})}$ in the limits $m \rightarrow 0$ and $m\rightarrow \infty$. In the large mass limit, the density fluctuates on very short timescales, so that the gravitational influence on the AIs averages to zero. In the small-mass limit, instead, the quasiparticles increase in both mass and size, eventually settling into configurations that, on experimental length scales, resemble the average energy density of dark matter at a fixed and nonzero dark matter fraction.

Using Eqs.~\eqref{eq:toy_ultralight_phase}-\eqref{eq:toy_ultralight_SNR} and assuming a $\mathrm{SNR} \sim 1$, the peak sensitivity occurs at 
\begin{equation}
m^{(\star)} \lesssim 10^{-17}~\mathrm{eV} \left (\frac{10^8~\mathrm{s}}{T_\mathrm{int}}\right )
\end{equation}
and dark matter fraction
\begin{equation}
\begin{aligned}
f_\mathrm{DM}^{(\star)} \sim\, & \mathcal{O}(\mathrm{10}) \,
\left( \frac{150\, {\rm s}}{T} \right)^{2} 
\left( \frac{250}{n} \right)
\left( \frac{10^8\,\mathrm{s}}{T_{\rm int}} \right) \\
& \times
\left( \frac{4.4 \times 10^7 \,{\rm m}}{L} \right)^{1/2} \left( \frac{S_n}{10^{-8} \, {\rm Hz}^{-1}} \right)^{1/2}  \, .
\end{aligned}
\end{equation}
Remarkably, proposed detectors could probe a DM overdensity of $\mathcal{O}(10)$ times the local DM energy density via purely gravitational interactions.  Moreover, AGs would measure the local DM density on a much shorter length scale compared to complementary probes. For instance, measurements using planetary ephemerides set an upper limit of $f_\mathrm{DM} \lesssim 10^4$~\cite{Pitjev:2013sfa}, while lunar laser ranging and the LAGEOS geodesic satellite set an upper bound of $f_\mathrm{DM} \lesssim 10^{11}$~\cite{Adler:2008rq}. The Doppler tracking data from the Cassini spacecraft~\cite{Bertotti:2003rm} excludes $f_\mathrm{DM}\gtrsim 10^6$ for $m\sim10^{-21}$~eV and DM overdensities larger than $\sim 10^{8}$ the local DM energy density for $m\sim10^{-19}$~eV~\cite{Blas:2016ddr}, while tracking of binary systems (e.g., millisecond pulsars, the Earth-Moon system via lunar laser ranging and Earth-satellite system via satellite laser ranging) may probe DM overdensities $f_\mathrm{DM}\gtrsim 10^3$ for $10^{-21}~\mathrm{eV} \lesssim m\lesssim10^{-19}$~eV~\cite{Foster:2025nzf}. PTAs probe $f_\mathrm{DM} \sim 1$ for $m\lesssim 10^{-23}$~eV and large dark overdensities up to $\sim 10^{-22}$~eV~\cite{NANOGrav:2023hvm}.

In light of these observations, in the following subsections we calculate the full AG phase shift signal using the formalism introduced in section~\ref{subsec:formalism}. As we show in the following subsections, the estimate provided here is in excellent agreement with the full calculation and can be used to understand the qualitative features of the reach plot. In particular, in Newtonian gauge, we find that the leading order contribution to the observable originates from the gravitational redshift measured at two distinct spacetime points, i.e. the Einstein phase shift.

\subsection{Full signal calculation}\label{sec:ultralight:signal}

For simplicity, let us assume that DM is a free massive real scalar field $\phi$, which is only minimally coupled to gravity. We model the scalar as
\begin{equation}
\phi(t,\mathbf{x}) = \phi_0 \cos(m t- \mathbf{k}\cdot\mathbf{x} + \theta) \, ,
\end{equation}
where $\phi_0$ is the slow-varying amplitude of the field, $\mathbf{k} = m \mathbf{v}$ is the momentum of the field and $\theta$ is a random phase. Since scalar DM can be understood as an incoherent superposition of Fourier modes with velocity $v$ sampled from the DM velocity distribution, $\phi$ is phase coherent within a coherence time $\tau_c = \sqrt{2\pi}\mathrm{Erf}[\sqrt{2}v_\mathrm{obs}/v_0]/mv_0v_\mathrm{obs}$ and coherence length $\lambda_c = \sqrt{2\pi}/mv_0$, where we used the convention recently introduced in Ref.~\cite{Cheong:2024ose}. Furthermore, in this model, the amplitude of the field is a Rayleigh-distributed random variable with $\langle \phi_0^2\rangle = 2 f_\mathrm{DM}\rho_\mathrm{DM}/m^2$, which is sampled every coherence length and coherence time.\footnote{Here, we neglect corrections to the amplitude of the DM field due to gravitational focusing, which would give rise to a $\lesssim \mathcal{O}(10\%)$ and time-varying correction to the dark matter energy density at a distance of 1 AU from the Sun~\cite{Kim:2021yyo}.}  

To determine the gravitational influence of this field on test masses, we compute the field's energy-momentum tensor $T_{\mu\nu}$ and then solve the linearised form of Einstein's equations. As we derive in Appendix~\ref{appendix:einstein_eqn}, in Newtonian gauge the leading order and fast oscillating nonzero entries of $h_{\mu\nu}$ read as
\begin{equation}\label{eq:metric_ULDM}
\begin{aligned}
h_{00} &= \pi G \phi_0^2 \cos(2m t- 2\mathbf{k}\cdot\mathbf{x} + 2\theta)  \\
h_{ij} &= -\pi G \phi_0^2 \cos(2m t- 2\mathbf{k}\cdot\mathbf{x} + 2\theta)\delta_{ij} \, .
\end{aligned}
\end{equation}
Using~Eqs.~\eqref{eq:total} and \eqref{eq:LMT_tot_F}, working in the limit in which the distance from the AI to the nearest laser is negligible with respect to the length of the baseline (i.e. setting $x_\mathrm{L_1} = x_\mathrm{AI_1}$ and $x_\mathrm{L_2} = x_\mathrm{AI_2}$ in Eq.~\eqref{eq:LMT_tot_F}) and assuming that $L/\lambda_c\ll 1$, the three contributions to the total atom gradiometer phase shift read as
\begin{widetext}
\begin{equation} \label{eq:ULDM_phase}
\begin{aligned}
\Delta \widetilde{\phi}^\mathrm{LMT}_\mathrm{grad,\mathcal{D}}(\omega) &=
\frac{1}{2}\, k_{\rm eff} \, T^2 \, \hat{\vec{n}} \cdot  \mathbf{k} \,
\left( K_+(\omega) - K_-(\omega) \right ) \left ( 1-e^{\mi \omega L}\right )  \widetilde{h}_{00}(\omega) \, , \\
 \Delta \widetilde{\phi}^\mathrm{LMT}_\mathrm{grad,\mathcal{E}}(\omega) & = 
- \frac{\mi }{4} \, k_{\rm eff} \, T^2 \, \omega \,
\left( K_+(\omega) + K_-(\omega) \right ) \left ( 1-e^{\mi \omega L}\right )  \widetilde{h}_{00}(\omega) \, , \\ 
\Delta \widetilde{\phi}^\mathrm{LMT}_\mathrm{grad,\mathcal{S}}(\omega) &= 
 0 \, .
\end{aligned}
\end{equation}
\end{widetext}
Notice that in this gauge the Shapiro gradiometer phase shift is exactly zero, which follows from the relative minus sign between the $h_{00}$ and $h_{ij}$ components of the metric tensor (cf. Eqs.~\eqref{eq:TScurl}--\eqref{eq:mathcal_H}). This may also be understood at the level of the action. Gravity couples to matter and radiation via the operator $h_{\mu\nu}T^{\mu\nu}$. In Newtonian gauge, we found that fluctuations in the energy density and pressure of ultralight dark matter give rise to spacetime fluctuations given by $h_{\mu\nu}=\Phi \eta_{\mu\nu}$, where $\Phi=-\pi G \phi_0^2 \cos(2m t- 2\mathbf{k}\cdot\mathbf{x} + 2\theta)$. Hence, $\Phi$ couples to test particles as $\Phi T^{\mu}_{\mu}$. Since the trace of the energy momentum tensor of electromagnetism is zero at the classical level and in the absence of external currents, $\Phi$ does not interact affect the free evolution of photons along the baseline. Furthermore, the Doppler term is $k$-suppressed, since the tidal displacement of the atomic clouds in the AG is DM velocity suppressed (cf. Eq.~\eqref{eq:tidal_displacement}). Additionally, the Doppler term depends on the combination of detector response functions $K_+-K_-$, while the Einstein term depends on the combination $K_++K_-$; consequently, in the limit $\omega n L, \omega T \ll 1$, the Doppler contribution receives a suppression $\sim 2L/T \ll 1$ relative to the Einstein contribution. Therefore, at leading order the signal amplitude is given by
\begin{equation}\label{eq:ULDM_signal}
\begin{aligned}
\sqrt{\langle |\Delta\phi^\mathrm{LMT}_\mathrm{grad}|^2\rangle} & =  \frac{4 \pi G \omega_a f_\mathrm{DM}\rho_\mathrm{DM}}{m^3}  \\ &  \qquad \times \sin \left ( m T \right ) \sin \left ( m n L \right )  \\ & \qquad \times \sin \left ( m (T-(n-1)L)\right ) \, .
\end{aligned}
\end{equation}
A reader familiar with GW and DM phenomenology at AGs may notice that Eq.~\eqref{eq:ULDM_signal} can be mapped directly to the signal induced by transient GWs~\cite{Graham:2016plp,Badurina:2024rpp} and linearly-coupled scalar ULDM~\cite{Arvanitaki:2016fyj,Badurina:2021lwr}. In the former case, the results match under the replacement $m\rightarrow \omega/2$ and $f_\mathrm{DM}\rightarrow \omega^2 h/(8\pi G \rho_\mathrm{DM})$, whilst in the latter case we have $m\rightarrow m/2$ and $f_\mathrm{DM}\rightarrow d_{m_e} m/\sqrt{2\pi G \rho_\mathrm{DM}}$, where $h$ is the strain and $d_{m_e}$ is the coupling to electron masses. This should come as no surprise: in the limit where the field is phase-coherent over the spatial extent of the AG, the induced gravitational signal is identical to that induced by a gravitational wave oscillating with angular frequency $\omega=2m$. The similarities between this effect and the signal induced by linearly-coupled scalar ULDM can be understood from the choice of gauge. In Newtonian gauge, we found that the leading-order gravitational contribution can be traced back to the Einstein phase shift term. This term accounts for the gravitational redshift measured by each atomic wavepacket, or equivalently the gravitational correction to the energy difference between the ground and excited states of the atom. The latter interpretation is exactly analogous to the effect induced by linearly-coupled scalar ULDM: any dilatonic coupling between a background scalar field and, e.g., electron masses and photons, induces a time-dependent correction to the energy difference between the two ground and excited states of the atom.

\subsection{Sensitivity analysis \& projected reach}\label{sec:ultralight:sensitivity}

To estimate the projected reach, we use matched filtering (cf. Eq.~\eqref{eqn:SNR}). However, unlike the DM clump case, here the SNR depends on the duration of the measurement campaign $T_\mathrm{int}$ relative to the coherence time $\tau_c$ of the field. Indeed, in the limit $T_\mathrm{int} \ll \tau_c$, the experiment cannot resolve the full spectral content of the field. Therefore, in this regime, the SNR increases with $\sqrt{T_\mathrm{int}}$. In contrast, in the regime where $T_\mathrm{int} \gg \tau_c$, the experiment fully resolves the full spectral content of the signal, whose bandwidth is $\sim 1/\tau_c$. In this regime, the SNR only grows as $(\tau_c T_\mathrm{int})^{1/4}$.

\begin{figure}[t!]	\includegraphics[width=0.48\textwidth]{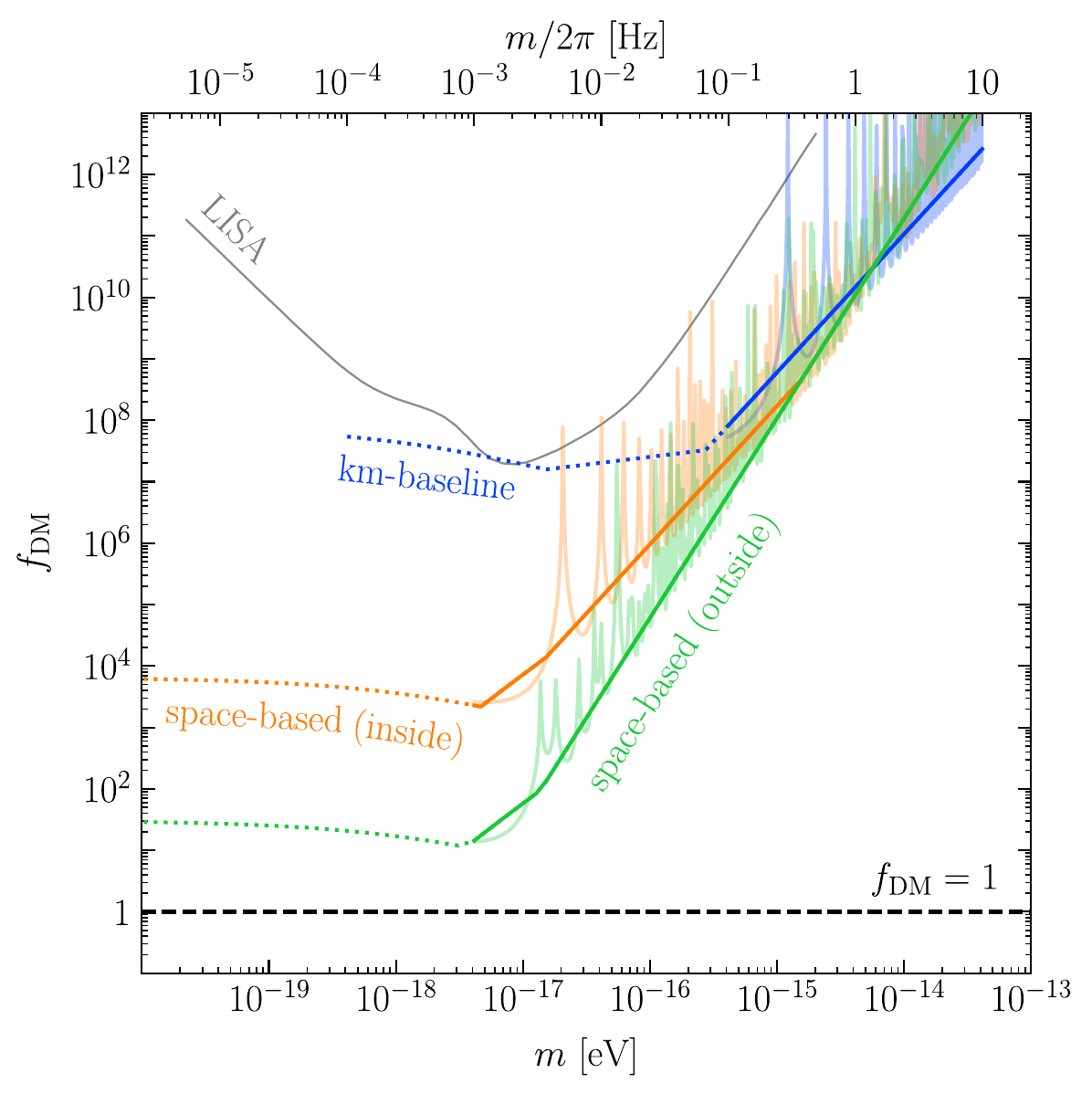} 
    \caption{Projected 90\% upper limits on the fraction of DM ($f_\mathrm{DM}$) composed of ULDM fields of mass $m$. We assume an experimental observation time $T_\mathrm{int}=10^8$~s and $\rho_{\mathrm{DM}}=0.46$~GeV/cm$^3$. We consider three (broadband) AG concepts: a terrestrial experiment with a km-long baseline (blue), a space-based experiment consisting of two AIs that are confined within two satellites in mid-Earth orbit (orange), and a space-based gradiometer that enables the free evolution of the atoms outside each satellite (green). The parameters for these experiments are summarized in Table~\ref{tab:exps}. For AGs we show power-averaged projections with darker lines. Solid and dotted lines show the reach in benchmark and optimistic scenarios. Projections for LISA, as derived in Ref.~\cite{Kim:2023pkx}, are also plotted (grey) for comparison.}
    \label{fig:ULDMreach_det}
\end{figure}

In Fig.~\ref{fig:ULDMreach_det} we plot the 90\% projected upper limits on the DM fraction as a function of the DM mass for the experiments listed in Table~\ref{tab:exps}. In light of ULDM's statistics, the 90\% upper limits on $f_\mathrm{DM}$ correspond to $\mathrm{SNR}_t \approx 11$ for $T_\mathrm{int} \ll \tau_c$ and $\mathrm{SNR}_t \approx 2.6$ for $T_\mathrm{int} \gg \tau_c$~\cite{Berlin:2020vrk}.\footnote{When $T_\mathrm{int}\ll \tau_c$, one samples a single value of the stochastically fluctuating and
Rayleigh-distributed DM field amplitude, which may be smaller than its long-time average; to account for this, the
exclusion SNR threshold is larger in the short integration time regime, as also discussed in Ref.~\cite{Centers:2019dyn}.}
As for the dark clump case, we plot the projected reach for benchmark and optimistic experimental bandwidths, which we display with solid and dashed lines, respectively. Furthermore, we show with darker lines the power-averaged envelope using the approximation $|\sin x| = 
\min (x, 1/ \sqrt{2})$, which smoothens the peaks and troughs arising from the fast oscillation of the detector response kernels in Eq.~\eqref{eq:ULDM_signal}. Note that the power-averaged sensitivity is in excellent agreement with our heuristic estimate. Due to the change in $\mathrm{SNR}_t$ in the long and short integration time limits, the sensitivity curves for space-based proposals plateau for $m\ll 1/T$, corresponding to regime (A) in Eq.~\eqref{eq:toy_scalings_ultralgiht}, and transition to regime (C) for $m\gtrsim 1/T$. For $m\gtrsim 1/nL$, the sensitivity curves exhibit the mass scaling of regime (D). The projected reach of the ``space-based (outside)" proposal transitions from regime (C) to (D) at smaller ULDM masses than the ``space-based (inside)" configuration because of the much larger number of LMT kicks (cf.~Table~\ref{tab:exps}). For the terrestrial experiment considered here, we also find that the curve plateaus at low masses after rising due to the increase in $\mathrm{SNR}_t$ in the $T_\mathrm{int} \ll \tau_c$ regime. However, at $m~\sim 10^{-17}$~eV, when the integration time exceeds the coherence time of the signal, the projected sensitivity exhibits a $m^{1/4}$ mass scaling, in agreement with the arguments presented in section~\ref{sec:ultralgiht_heuristics}.

As anticipated in section~\ref{sec:ultralgiht_heuristics}, space-based gradiometers may be able to probe a DM overdensity of $\mathcal{O}(\mathrm{10})$ times the local DM energy density for masses below $m\lesssim 10^{-17}$~eV. 
Furthermore, we find that the space-based AG proposals would be able to set much stronger constraints on dark matter overdensities than other interferometric probes.
Indeed, as we show in Fig.~\ref{fig:ULDMreach_det}, LISA is projected to only probe DM fractions many orders of magnitude greater than unity, and proposed laser interferometer experiments such as BBO and $\mu$Ares would also set much weaker constraints~\cite{Kim:2023pkx}.

The remarkable reach of space-based AGs arises from their strain sensitivity and measurement type. In the limit $\omega L \ll 1$, the square root of the strain noise PSD is approximately given by $\sqrt{S_a(\omega)}/\omega^2 L$, where $\sqrt{S_a(\omega)}$ is the AG acceleration noise defined in Eq.~\eqref{eq:S_a}. Therefore, the ratio of the LISA and AG strain noise spectra is given by Eq.~\eqref{eq:ratio_spectra} times the ratio of the AG and LISA baseline lengths, which is approximately 1/50. In light of this correction, the strain sensitivity of LISA is comparable to that of the ``space-based (inside)" proposal, while the ``space-based (outside)" design is more sensitive by a few orders of magnitude. Furthermore, AGs can be understood as
clock comparison experiments~\cite{Dimopoulos:2006nk,Dimopoulos:2007cj,Dimopoulos:2008hx}, due to the dependence of the
observable on the Einstein time delay measured by the two AIs~\cite{Badurina:2024rpp}. 
This differs from the measurement performed by a two-arm laser interferometer, such as LISA and LIGO~\cite{Lee:2024oxo}. In laser interferometers and in Newtonian gauge, the fast-oscillating fluctuations of the ULDM's pressure and energy density exclusively influence the tidal displacement of the mirrors (i.e. Doppler time delay), which is $v$-suppressed. Indeed, as shown in Ref.~\cite{Kim:2023pkx}, in the relevant mass range, the signal in laser interferometers is suppressed by $L/\lambda_c \ll 1$. Hence, for any value of $m$, an AG's sensitivity to $f_\mathrm{DM}$ is expected to be parametrically enhanced with respect to that of a laser interferometer with comparable strain sensitivity.

As for the dark clump case, note that the projections for the ``space-based (outside)'' proposal are based on noise curves that neglect astrophysical foregrounds. In light of Figs.~\ref{fig:ULDMreach_det} and \ref{app:fig:noise}, we anticipate that an AEDGE$+$-like detector would still achieve stronger (though comparable) constraints on $f_\mathrm{DM}$ than the ``space-based (inside)'' proposal. In particular, we expect this setup to probe $f_\mathrm{DM} \gtrsim 10^3$ for $m\lesssim 10^{-16}$~eV. Due to uncertainties in the modeling of astrophysical foregrounds, especially in the $10^{-3}$--$10^{-2}$~Hz band (see Appendix~\ref{app:noise_sources}), our benchmark scenario may still capture the expected reach, in which case values of $f_\mathrm{DM} \gtrsim 10$ could still be accessible for $m \sim 10^{-17}$~eV.

We finally note that PTAs would also be sensitive to the same parametric enhancement, since the time-residual observable also depends on a variant of the Einstein time delay~\cite{Maggiore:2007ulw}. Furthermore, the deterministic signal due to ULDM fluctuations peaks at $m\lesssim 1/L$, where $L$ is the typical Earth-pulsar distance.
However, differently from the AG case, the reach is limited by the experimental bandwidth. Since $\omega_\mathrm{min}\gg 1/L$, the sensitivity curves exhibit the mass scaling of regime (D) in Eq.~\eqref{eq:toy_scalings_ultralgiht} and the peak sensitivity occurs at $m\sim 10^{-23}$~eV~\cite{Khmelnitsky:2013lxt,Kim:2023kyy}.

\section{Discussion \& Conclusions}\label{sec:discussion_conclusion}

In this work, we explored the prospects of detecting the purely gravitational signatures of DM from the ultralight to the ultraheavy mass range with proposed atom gradiometer experiments. Using the formalism developed in our previous work~\cite{Badurina:2024rpp}, we derived the projected upper limits on the dark matter fraction saturated by compact DM clumps and minimally-coupled ULDM for several detector concepts, which are based on MAGIS-km, AION-km, AEDGE and AEDGE+. Although terrestrial experiments are unlikely to probe interesting regions of parameter space, our findings suggest that, under optimistic noise projections, a space-based AG akin to AEDGE+ may be able to probe a $f_\mathrm{DM}\sim \mathcal{O}(0.1)$ DM fraction saturated by clumps of mass $10^6\,\mathrm{kg} \lesssim M \lesssim 10^{10}\,\mathrm{kg}$ ($10^{-25}~M_\odot\lesssim M \lesssim 10^{-21}~M_\odot$), and a $f_\mathrm{DM}\gtrsim 10$ ULDM overdensity for masses $m \lesssim 10^{-17}$~eV. The parametrically enhanced reach of AGs relative to laser interferometers, such as LIGO and LISA, is due to: (\textit{i}) the exquisite acceleration sensitivity of these devices, and (\textit{ii}) the nature of the gravitational observable in AGs, which also depends on the relative gravitational redshift measured by the freely falling atoms in each AI. In light of (\textit{i}), AGs have parametrically enhanced sensitivity to gravitational signals induced by dark clumps. In light of (\textit{ii}), the sensitivity of an AG to the fast-oscillating component of the metric perturbation sourced by ultralight dark matter fluctuations is parametrically enhanced. 

In light of the remarkable reach of these ambitious proposals, there is scope to go beyond the analysis presented in this work. For instance, a detailed noise analysis that extends the discussion in Appendix~\ref{app:noise_sources}, especially for space-based proposals, is warranted to ($i$) precisely determine the impact of astrophysical backgrounds on the sensitivity of these experiments and ($ii$) quantify the instrumental noise budget well below $\sim 10^{-3}$~Hz, which AG noise analyses have neglected thus far. Although we partially accounted for this uncertainty with different values of $\omega_\mathrm{min}$, a more refined analysis is a natural extension of this work.

Additionally, we exclusively focused on broadband detection strategies. However, it is expected that GW searches with very long baselines will also operate in resonant mode~\cite{Graham:2016plp}: in these configurations, the peak response to a transient GW occurs at $\omega\sim 1/T$; the signal is amplified by the quality factor $Q$, but its bandwidth is reduced by $Q$. The interrogation time and the number of LMT kicks are bounded by $2QT\leq T_\mathrm{max}$ and $2Qn\lesssim n_\mathrm{max}$, respectively, where $T_\mathrm{max}$ and $n_\mathrm{max}$ are dictated by experimental considerations. Since resonant mode detection strategies may be beneficial in searching for gravitational waves with frequencies $\omega \gg 1/T$, we leave a careful treatment of different scanning strategies for future work. 

Furthermore, we focused on a subset of time-dependent gravitational signals due to dark clumps and ULDM. Specifically, we studied the deterministic signal caused by a single transient clump, and the almost monochromatic ULDM signal peaked at $\sim 2m$. However, both DM candidates also give rise to stochastic signals. For fixed $f_\mathrm{DM}$ and in the small dark clump mass limit, an individual clump may not give rise to a detectable signal, but the collective effect due to the passage of multiple clumps in the vicinity of the detector may be large
enough to become observable. In this limit, the total signal is stochastic, as studied in the context of PTAs, e.g. Ref.~\cite{Ramani:2020hdo}. In the ULDM case, instead, the stochastic signal is associated with the slow-varying component of the energy density and pressure fluctuations that source metric perturbations. As discussed in Ref.~\cite{Kim:2023pkx}, the signal has support between zero and $\sim m v^2$, where $v \sim 10^{-3}$ is the mean DM speed, and is not velocity suppressed (since the perturbations enter at zeroth order in the DM velocity). Because this signal is analogous to that of a stochastic gravitational wave background, a search that cross-correlates the output of a network of spatially separated detectors would be necessary to disentangle the signal from colored noise. A study that considers these stochastic signals is left for future work.

Furthermore, we focused on a subset of time-dependent gravitational signals. For example, unlike the stochastic signal induced by DM clumps, the stochastic signal associated with the slow varying part of the spacetime fluctuations induced by ULDM is not expected to be parametrically suppressed~\cite{Kim:2023pkx}. Since this signal is analogous to that of a stochastic gravitational wave background, a search that cross-correlates the output of a network of spatially separated detectors would be ideal. A study in this direction, which would make use of multigradiometer configurations~\cite{Badurina:2022ngn}, is left for future work. 

In this study, we have also been model-agnostic in order to render our results as general and applicable as possible. It would therefore be natural to extend our work by considering concrete DM models. In the context of DM clumps, for example, PBHs of such a small mass would have evaporated by now and are therefore not viable candidates in the mass range of interest~\cite{Green:2020jor}.  Consequently, it would interesting to explore the role of a clump's compactness on the gravitational signal, which could then be used to study the gravitational signatures of less compact transient structures, e.g., axion and dark photon stars~\cite{Gorghetto:2022sue,Gorghetto:2024vnp}, or dark matter substructure more generally~\cite{Lee:2020wfn,Ramani:2020hdo}. Additionally, it would be interesting to explore the interplay between gravitational and non-gravitational signals. For example, many DM models include additional interactions, such as long-range Yukawa forces mediated by ultralight bosons, which could contribute to the phase shift measured by an atom gradiometer. 
For the ULDM case, it would be natural to extend our discussion to include DM self-interactions. In the case of an ultralight scalar, we would naturally expect the theory to exhibit a discrete shift symmetry, which would appear in the scalar potential through a cosine term, and would therefore give rise to a series of self-interactions satisfying a $\mathrm{Z}_2$ symmetry, i.e. $m^2 \phi^{2n}/\Lambda^{2n-2}$, where $n\geq 2$ and $\Lambda \gg m$. Since these quartic self-interactions naturally give rise to ULDM overdensities in our solar system~\cite{Budker:2023sex}, we leave a study on the impact of these self-interactions on the expected AG signal for future work.

\begin{acknowledgments}
L.B. would like to thank Marek Lewicki and Ville Vaskonen for helpful discussions on the impact of astrophysical foregrounds on the projected reach of space-based AG proposals, and Federico Cima for carefully reading the manuscript. L.B., Y.D., V.L., Y.W. and K.Z. are supported by the U.S. Department of Energy, Office of Science, Office of High Energy Physics under Award Number DE-SC0011632, and by the Walter Burke Institute for Theoretical Physics. VL is supported by NSF Physics Frontier Center Award \#2020275 and the Heising-Simons Foundation. Y.W. is also supported by~Grant 147323~Award \#704089 from~Fermi~National~Accelerator~Laboratory. KZ is also supported by a Simons Investigator award, and by Heising-Simons Foundation “Observational Signatures of Quantum Gravity” collaboration grant 2021-2817. The computations presented here were conducted in the Resnick High Performance Computing Center, a facility supported by Resnick Sustainability Institute at the California Institute of Technology.
\end{acknowledgments}

\appendix

\section{Noise sources} \label{app:noise_sources}

In this appendix, we justify the choice of $\omega_\mathrm{min}$ for the detector concepts considered in this work.

For Earth-based detectors, it is anticipated that these devices will be atom shot-noise limited. As discussed in Ref.~\cite{MAGIS-100:2021etm}, other sources of noise, e.g., laser phase noise, etc., are expected to be subdominant above $\sim 10^{-4}$~Hz for a device such as MAGIS-km, which is anticipated to operate with parameters comparable to those considered in this work. As discussed in Refs.~\cite{Junca_2019,Mitchell_2022_err,Badurina:2022ngn,Carlton:2024lqy}, however, the sensitivity of all terrestrial experiments will be seriously limited by gravity gradient noise (GGN) due to mass-density fluctuations in the vicinity of the experiment. These can arise as a result of atmospheric events and seismic activity, which limit the projected reach of the experiment in the sub-Hz regime. The spatial dependence of these noise contributions, as well as the possibility of using auxiliary sensors to perform complementary measurements, suggest that it may be possible to mitigate these backgrounds via active and passive noise mitigation/cancellation strategies, e.g., multigradiometer configurations~\cite{Badurina:2022ngn, Mitchell_2022_err} or Wiener filters based on correlations between the datasets of auxiliary sensors and AGs~\cite{2000rdgr.conf..495C,Canuel:2019abg}. Since the magnitude of these noise contributions significantly depends on the experiment's environment (e.g., geology, climate, etc.), we choose $\omega_\mathrm{min}/2\pi \in \{10^{-4}~\mathrm{Hz},10^{-1}~\mathrm{Hz}\}$.

\begin{figure}[t!]
    \centering
    \includegraphics[width=0.48\textwidth]{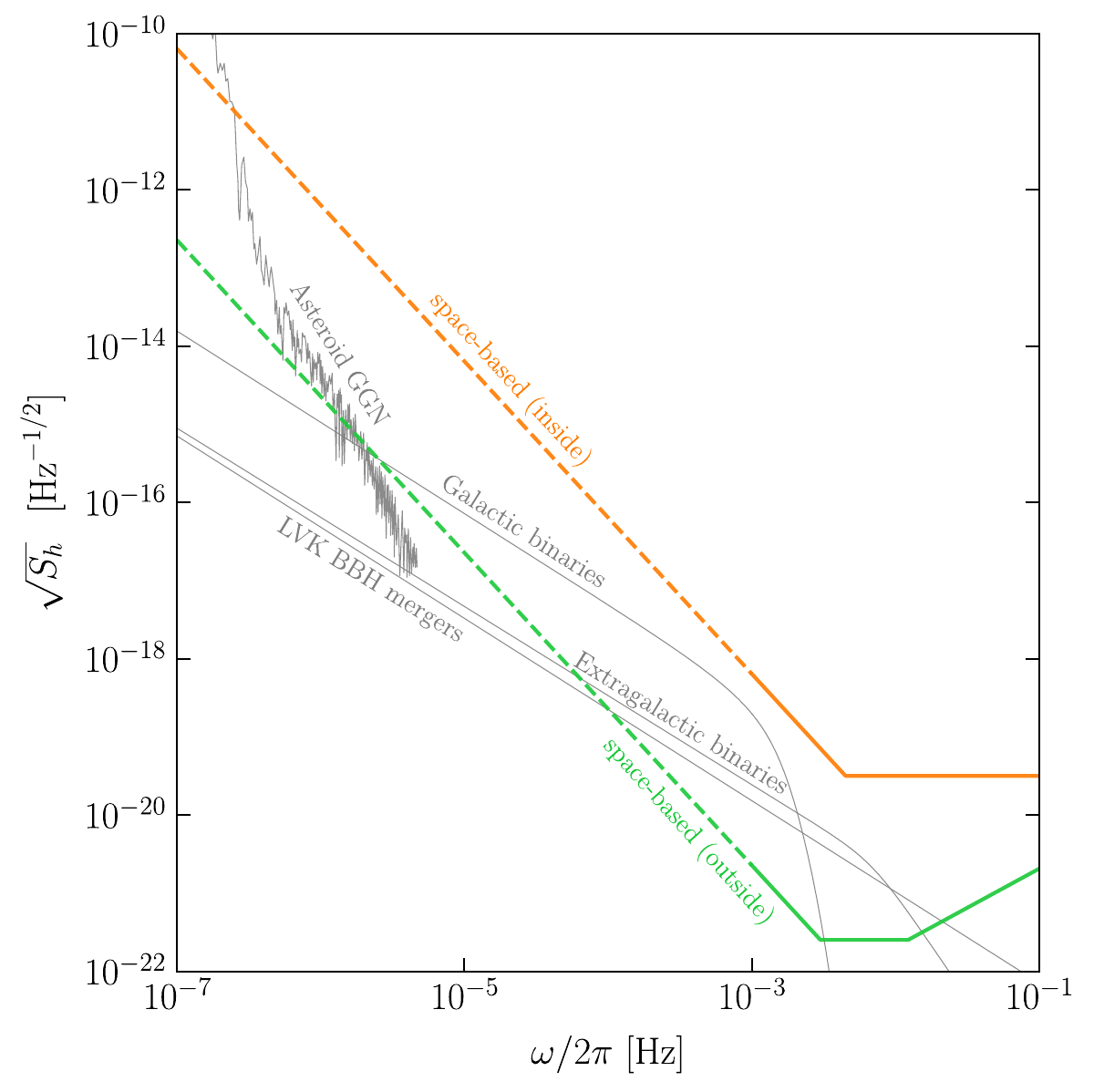}
    \caption{Power-averaged atom shot-noise limited strain sensitivity of the ``space-based (inside)" (orange) and ``space-based (outside)" (green) proposals (cf. Table~\ref{tab:exps}).   Astrophysical backgrounds due to asteroid gravity gradient noise, unresolved galactic and extragalactic white dwarf binary mergers, as well as LIGO-Virgo-KAGRA binary black hole (LVK BBH) mergers, are shown with gray curves. We show the extrapolated atom shot-noise limited strain sensitivity below $10^{-3}$~Hz with dashed lines.}
    \label{app:fig:noise}
\end{figure}

For both space-based proposals, the instrumental noise budget is anticipated to be shot-noise limited at least above $10^{-3}$~Hz, as discussed in Ref.~\cite{Dimopoulos:2008sv}. However, the ``space-based (outside)" and ``space-based (inside)" proposals are expected to be sensitive to astrophysical backgrounds below $\sim 10^{-2}$~Hz and $\sim 10^{-6}$~Hz, respectively. This can be understood from the strain sensitivity of the two experiments. As shown in Fig.~\ref{app:fig:noise}, GGN from asteroids as estimated in Ref.~\cite{Fedderke:2020yfy} is expected to become detectable below $10^{-7}-10^{-6}$~Hz. Unlike the other designs, the ``space-based (outside)" configuration would be able to detect the astrophysical foreground from unresolved galactic~\cite{Karnesis:2021tsh} and extragalactic white dwarf binary mergers~\cite{Farmer:2003pa,Staelens:2023xjn,Hofman:2024xar,Boileau:2025jkv}, as well as the foreground of binary black hole (BBH) mergers detectable by LIGO-Virgo-KAGRA (LVK)~\cite{Lewicki:2021kmu}. Using the parameters in Table~\ref{tab:exps}, we find that the background due to unresolved galactic binaries exceeds the noise curve of this experimental concept below $\sim 10^{-3}$~Hz. The background from unresolved extragalactic binaries, instead, dominates in the $10^{-3}-10^{-2}$~Hz frequency band. 
Note, however, that this background is largely model dependent, and there exist strategies to mitigate similar backgrounds due to neutron star binaries in this frequency window, e.g. Ref.~\cite{Yagi:2011wg}. Therefore, in this work we take $\omega_\mathrm{min}/2\pi \in \{10^{-6}~\mathrm{Hz},10^{-3}~\mathrm{Hz}\}$ for both experiments and leave a detailed study to future work.

\section{Minimum impact parameter}\label{app:min_impact_parameter}

In this appendix, we provide a detailed analysis of the minimum impact parameter used in deriving the projected reach in section~\ref{sec:ultraheavy:sensitivity}. The local statistical distribution of the impact parameter from a compact DM clump has been studied and derived in the appendix of Ref.~\cite{Dror:2019twh}, which we summarize here. 

For the Doppler and Einstein gradiometer phase shift, the relevant distance scale is $b$, since the DM effect acts only on a point (the atoms). Since events with fixed $|b|$ at one of the two AIs imprint the same phase shift, the statistics of $b$ are derived by randomly placing clumps in a (fictitious) cylindrical volume, with its height set by the average distance traveled by the clump, namely $\bar{v}T_{\rm int}$, where $\bar{v}$ is the DM average speed and $T_{\rm int}$ is the total experiment (or integration) time, and the cross-sectional area, $\pi R^2$, which is taken to infinity. The total number of DM clumps within this volume is given by $N=\pi R^2 \bar{v}T n_\mathrm{DM}$, where $n_{\mathrm{DM}}=\rho_{\mathrm{DM}}f_{\mathrm{DM}}/M$ is the local clump number density, $\rho_{\mathrm{DM}}=0.46\,\text{GeV}/\text{cm}^3$ is the local DM energy density, and $f_{\mathrm{DM}}$ is the DM fraction. The cumulative distribution function (CDF) of the impact parameter of each clump within the cylindrical volume is then given by $F(b)=(b/R)^2$. For notational convenience, let each clump be labeled by $j \in \{1,...,N\}$. Noting that the impact parameter of each clump is independent and identically distributed (i.i.d.), the CDF of the \textit{minimal} impact parameter is thus given by
\begin{align}\label{eqn:minimal_parameter}
    F_{{\min}}(b_{\min}) &= 1-P(b_{\min}<b_j\,\mathrm{for\,all\,}j) \nonumber \\
    &= 1-\left[1-F(b_{\min})\right]^N \nonumber \\
    &= 1-\left[1-\left(\frac{b_{\min}}{R}\right)^2\right]^N \nonumber \\
    &\to 1-e^{-\pi \bar{v}T_{\rm int}n_{\mathrm{DM}}b_{\min}^2} \, ,
\end{align}
where we took the large $N$ limit in the last line. Importantly, we see that both the fictitious parameters $R$ and $N$ drop out of the CDF in the $N\to \infty$ limit. The $p^{\mathrm{th}}$ percentile minimum value is then defined by setting $F_{{\min}}(b_{\min})=p$ and solving Eq.~\eqref{eqn:minimal_parameter} for $b_{\min}$, i.e.
\begin{equation}\label{eqn:b_min}
	 b_{\min} = \sqrt{-\frac{\ln(1-p)}{\pi \, n_{\mathrm{DM}}\, \bar{v}\,T_{\rm int}}}
\end{equation}
Setting $p = 0.9$, we infer the $90^\mathrm{th}$-percentile of the minimal impact parameter.

For the Shapiro gradiometer phase shift, we consider two regimes. If the DM is sufficiently distant ($\gtrsim L/2$) from the detector, the entire interferometer arm is effectively a point and the relevant impact parameter is still $b$, so that the analysis for the Doppler and Einstein terms follows through (i.e., Eqs.~\eqref{eqn:minimal_parameter}-\eqref{eqn:b_min}). However, for nearby clumps ($b\lesssim L/2$), the relevant length scale for the Shapiro contribution is the DM's closest encounter to any point along the baseline, which we denoted in section~\ref{sec:ultraheavy:signal} as $b^\pm_{\perp}$ (note that $b_{\perp}^\pm\leq b$ by definition). In this case, the geometry of the volume around the AG is rectangular, with width $L$, length $R$, and height $\bar{v}^\pm_{\perp}T_{\mathrm{int}}$, where $\bar{v}^\pm_{\perp}$ is (to leading order) the mean perpendicular speed of DM relative to the baseline. The total number of clumps inside the volume is thus $N=n_{\mathrm{DM}}\bar{v}^\pm_{\perp}T_{\mathrm{int}}LR$. The CDF of an individual DM impact parameter is given by $F(b^\pm_{\perp})=b^\pm_{\perp}/R$. Following the logic of Eq.~\eqref{eqn:minimal_parameter}, one can derive the $p^{\mathrm{th}}$ percentile minimum value of $b^\pm_{\perp}$ as
\begin{equation}
\label{eq:b_perp_min}
	 b^\pm_{\perp,\min} = -\frac{\ln(1-p)}{n_{\mathrm{DM}} \, \bar{v}^\pm_{\perp}\,T_{\rm int}\,L} \, .
\end{equation}

\section{Metric fluctuations sourced by ultralight dark matter}\label{appendix:einstein_eqn}

In this appendix, we explicitly compute the influence of ULDM's energy-density and pressure fluctuations on spacetime. Although we assume that ULDM is a scalar, our results also apply to the case where the field is a massive spin-1 particle. 

In flat spacetime, i.e. $\eta_{\mu\nu}=\mathrm{diag}(-1,1,1,1)$, a free scalar field satisfies the Klein--Gordon equation, $\Box \phi = m^2 \phi$. In the Standard Halo model, dark matter is assumed to be virialised. Consequently, the scalar field's amplitude and phase vary over time and length scales greater than the field's coherence time $\tau_c = 1/mv_0^2$ and coherence length $\lambda_c = 1/mv_0$, respectively. Since $\phi$ is non-relativistic, the field may be expressed as a fluctuating plane
wave, i.e.
\begin{equation}
\phi(t,\mathbf{x}) = \phi_0(t,\mathbf{x}) \cos(\omega t+\theta(t,\mathbf{x})) \, .
\end{equation}
Here, $\omega = m+mv^2/2$, $\phi_0(t, \mathbf{x})$ is the slow-varying amplitude of classical wave and $\theta(t, \mathbf{x})$ is the slow-varying component of its phase. We define the momentum of the field as $\mathbf{k} = \nabla \theta$, with $\partial_t \phi_0/\phi_0 \ll \omega$, $|\nabla \phi_0|/\phi_0 \ll k$ and $\partial_t \theta \ll \omega$.

The energy--momentum tensor of a free real scalar field takes the form
\begin{equation}\label{app:eq:T}
T_{\mu\nu} = \partial_\mu \phi \partial_\nu \phi - \frac{1}{2} \eta_{\mu\nu} \left ( \eta^{\alpha \beta} \partial_\alpha \phi \partial_\beta \phi + m^2 \phi^2 \right ) \, ,
\end{equation}
which we rewrite in terms of time-averaged $\left \langle T_{\mu\nu} \right \rangle$ and fast-oscillating $\delta T_{\mu\nu}$ components, i.e.
\begin{equation}
T_{\mu\nu} = \left \langle T_{\mu\nu} \right \rangle + \delta T_{\mu\nu} \, .
\end{equation}
By defining the energy density as $\rho + \delta \rho = T_{00}$ and the pressure as $P + \delta P =  T^i_{i}/3$ and neglecting slow-varying contributions, $\delta T_{\mu\nu}$ may be rewritten as
\begin{align}
    \delta \rho &=  -\frac{ \phi_0^2 }{2}k^2\cos(2\omega t-2\mathbf{k}\cdot\mathbf{x}+\theta) \, , \\
    \delta P &= -\frac{ \phi_0^2 }{6}(3m^2+k^2)\cos(2\omega t-2\mathbf{k}\cdot\mathbf{x}+\theta) \, , \\
    \delta T_{ij} &= -\frac{ \phi_0^2 }{2}k_ik_j\cos(2\omega t-2\mathbf{k}\cdot\mathbf{x}+\theta) \, , \, \text{for $i\neq j$} \, , \\
    \delta T_{0i} &=\frac{ \phi_0^2 }{2}\omega k_i\cos(2\omega t-2\mathbf{k}\cdot\mathbf{x}+\theta) \, .
\end{align}
Crucially, note that the fluctuating part of the energy density is of the same order as $\delta T_{ij}$.

Working in Newtonian gauge, i.e.,
\begin{equation}
g_{\mu\nu} = \mathrm{diag}(-1-2\Phi,1-2\Psi,1-2\Psi,1-2\Psi) \, ,
\end{equation}
the Einstein equations take the simple form~\cite{Maggiore:2018sht}
\begin{equation}
\begin{gathered}
\nabla^2 \Psi = 4\pi G \rho \,  , \\
\partial_t^2 \Psi + \frac{1}{3}\nabla^2(\Phi-\Psi) = 4\pi G P \, .
\end{gathered}
\end{equation}
Focusing exclusively on the Fourier modes with frequency components $\approx 2\omega$, we find 
\begin{equation}
\begin{aligned}
\Phi &= -\frac{\pi}{2} G \phi^2_0 \cos(2\omega t-2\mathbf{k}\cdot\mathbf{x}+\theta) \, , \\
\Psi &= \frac{\pi}{2} G \phi^2_0 \cos(2\omega t-2\mathbf{k}\cdot\mathbf{x}+\theta) \, .
\end{aligned}
\end{equation}
The nonzero entries of the perturbed metric tensor then read as
\begin{equation}
\begin{aligned}
h_{00} &= \pi  G \phi_0^2 \cos(2\omega t- 2\mathbf{k}\cdot\mathbf{x} + 2\theta) \, ,  \\
h_{ij} &= -\pi G \phi_0^2 \cos(2\omega t- 2\mathbf{k}\cdot\mathbf{x} + 2\theta)\delta_{ij} \, .
\end{aligned}
\end{equation}
Approximating $\omega$ as $m$, we recover the result stated in section~\ref{sec:ultralight:signal}, which is in agreement with Refs.~\cite{Dror:2024con,Kim:2023pkx}.

\bibliography{bib}{}
\bibliographystyle{apsrev4-2}

\end{document}